\documentclass[iop]{emulateapj}

\usepackage{natbib}
\usepackage{verbatim}
\usepackage{graphicx}
\usepackage{ulem}
\usepackage{cancel}
\usepackage{enumerate}
\usepackage{enumitem}

\begin{document}

\title
{Properties of spectrally-defined red QSOs at $z = 0.3-1.2$}

\author{
A.-L. Tsai\altaffilmark{1}
and C.-Y. Hwang\altaffilmark{1}}
\email{altsai@astro.ncu.edu.tw, hwangcy@astro.ncu.edu.tw}
\affil{
Institute of Astronomy, National Central University, 
No. 300, Jhongda Rd., Jhongli, Taoyuan 32001, Taiwan}

\begin{abstract}
We investigated the properties of a sample of red Quasi-stellar Objects (QSOs)
using optical, radio, and infrared data.
These QSOs were selected from
the Sloan Digital Sky Survey Data Release 7 (SDSS DR7) quasar catalog. 
We only selected sources with sky coverage of
the Very Large Array Faint Images of the Radio Sky at Twenty-centimeters (VLA FIRST) survey, 
and searched for sources with Wide-field Infrared Survey Explorer (WISE) counterparts.
We defined the spectral color of the QSOs 
based on the flux ratio of the rest frame
4000\AA~to 3000\AA~continuum emission
 to select red QSOs and typical QSOs.
In accordance with this criterion,
only QSOs with redshifts between 0.3 and 1.2 could be selected.
 We found that the red QSOs
 have stronger infrared emission than the typical QSOs do.
 We noted that the number ratios of the red QSOs to the typical QSOs
 decrease with increasing redshifts,
 although the number of the typical QSOs increase with redshifts.
Furthermore, at high redshifts,
 the luminosity distributions of the typical QSOs 
 and the red QSOs seem to have similar luminosity distribution peaks;
however, at low redshifts, the luminosities of the red QSOs
  seem to be lower than those of the typical QSOs.
These findings suggest that there might be at least two types of red QSOs in our QSO samples.

\end{abstract}


\maketitle
\thispagestyle{empty}

\section{Introduction}
Quasi-Stellar Objects (QSOs) are one type of active galactic nuclei (AGNs),
 and are among the most luminous objects in the universe.
The energy of AGNs is believed to be powered by 
 surrounding accretion disks feeding matter into
 a supermassive black hole (SMBH)  at the center of the host galaxy.
The accretion disk and SMBH are surrounded by a dusty torus
 that can obscure the line of sight in certain directions.
Some AGNs also have strong radio emission with relativistic jets perpendicular to the accretion disk
\citep{urr95}.

A typical QSO spectrum shows a non-stellar continuum with 
 strong broad line emission and narrow line emission.
In the rest frame,
 the continuum spectrum usually peaks at ultraviolet (UV) to soft X-ray wavebands. 
However, recent observations have discovered a new population of QSOs,
 which show redder optical or infrared (IR) colors than typical QSOs do
\citep{web95,ric03,gli07}.
No standard definition for these red QSOs currently exists.

The nature of these red QSOs is still unclear.
The redness could be attributed to several different effects,
 such as dust-reddening \citep{gli04,urr09}, 
 an intrinsic red continuum of these QSOs \citep{ric03}, 
 contamination from stars in host elliptical galaxies \citep{mas98},
 or additional red synchrotron emission \citep{whi01}.
The majority of the red QSOs discovered are considered
 to be reddened by dust \citep{cut02,gli04,can06}.
The origins of the red color could be explained by one of the following two scenarios: 
 (1) 
the lines of sight to the dust-reddened QSOs pass through the AGN torus;
 or
 (2) the dust is produced during starburst activities
 which follow the galaxy mergers that trigger QSOs.
The second scenario is likely related to an early evolutionary stage of QSOs.
However, 
 although dust might be a crucial effect of reddening, 
 we do not know  whether the reddening is caused entirely by dust.
Therefore, using various methods to test the potential effects 
 involved in reddening is essential.

The existence of dust in QSOs can significantly affect
 the observational properties of the QSOs and their host galaxies \citep{rud84}.
In optical observations, 
 a population of dust-obscured QSOs would be missed
 by magnitude-limited surveys.
The percentage of the missing QSOs
 might be from a few percent to $\sim30\%$ \citep{whi01}.
However, the optically obscured photons 
 re-emerge as emission at  the infrared band.
Therefore, the infrared emission might reveal properties of dust-obscured QSOs.
In addition, we noted that radio emission is barely affected by dust absorption,
 and that radio emission of the QSOs should be
 independent of the amount of dust in the sources.
To investigate the origins of the red color of QSOs, 
 we used optical, infrared, and radio data 
 to study the properties of selected QSO samples.

Most selection methods for finding red QSOs rely on photometric selection.
For example, 
 the parent QSO samples were selected from either optical \citep{ric03},
 infrared--infrared matching \citep{ban13},
 optical--infrared matching \citep{cut01,geo09,fyn13,ros15}, or
 optical--infrared--radio matching sources \citep{gli07,urr09,gli12}.
The criteria for selecting red QSOs were mainly based on
  their photometric colors such as
 optical--optical
 (e.g., $r-i$, $g-r$; \citealt{fyn13}),
 optical--infrared
 (e.g., $R-K$, $g-J$, $i-K$, $r-$W4; \citealt{cut01,gli07,geo09,urr09,gli12,fyn13,ros15}),
 and infrared--infrared colors 
 (e.g., $J-K_{\rm s}$, $J-K$, W1--W2; \citealt{cut01,gli07,geo09,urr09,gli12,ban13,fyn13}).
However, even for the same photometric colors,
 the red QSOs could be defined by different color criteria
 in different surveys;
 for example, the red QSOs were selected with
 $J-K>2$ and $R-K>5$~from 2MASS-SDSS samples by \citet{geo09},
 $J-K>1.7$ and $R-K>4$~from FIRST-2MASS samples by 
 \citeauthor{gli07} (\citeyear{gli07}, \citeyear{gli12}),
 and
 $J-K>1.3$ and $R-K>5$~from FIRST-2MASS samples by \citet{urr09}.
This implies that the red QSOs discussed
 in various studies might not be uniform.
Furthermore, other concerns of photometric selection include:
 redshifted strong emission lines that could contaminate the photometric colors,
and
 photometric data of the same filter representing
 different waveband properties at different redshifts.
To avoid these problems of photometric selection for red QSOs,
 \citet{ric03} defined red QSOs based on ``relative photometric colors'',
 which are the differences between
 the measured photometric colors and 
 the median colors of QSOs at the redshift of the QSO.
In this study, 
 we developed a new method for classifying red QSOs with
 a statistical definition
 based on ``relative spectral flux'' to not only facilitate avoidance of
 the inconsistency caused by photometric cut-offs,
 but also enable selection of statistically-defined red QSOs 
 to study the origins of their redness.
Since the redness of our QSOs was uniformly defined
 at the same rest-frame wavelengths over the whole redshift range we considered, 
 conducting follow-up confirmations for the reality of the redness is not necessary.

In this study, 
 we selected the parent QSOs from the
 Sloan Digital Sky Survey Data Release 7 (SDSS DR7) Quasar Catalog
 based on their spectra at the rest frame.
The SDSS DR7 QSOs included both photometrically selected and FIRST-matched QSO samples \citep{ric02}.
We chose QSOs with radio counterparts from
 the Very Large Array Faint Images of the Radio Sky at Twenty-centimeters 
 (VLA FIRST) radio survey, as well as 
 infrared counterparts from the Wide-field Infrared Survey Explorer (WISE).
We defined QSO colors based on the relative flux ratio from the spectrum.
Through this method,
 we could select red QSOs without the contamination of strong emission lines,
 thereby  enabling us to include a wide range of red QSOs
  to study their general properties.
The detail of the data selection is described in Section 2.
We present our results in Section 3 and 
 discuss the implication of our results in Section 4.
The cosmology parameters we are using in this paper are
 $H_{0} = 70$ km~s$^{-1}$~Mpc$^{-1}$,
  $\Omega_{\rm m} = 0.30$, and  $\Omega_{\Lambda} = 0.70$.

\section{Sample Selection}
\subsection{\rm Optical Data and Definition of QSO Color} 
\begin{figure}
\begin{center}
\begin{tabular}{cc}
{\includegraphics[height=8.5cm,angle=-90]{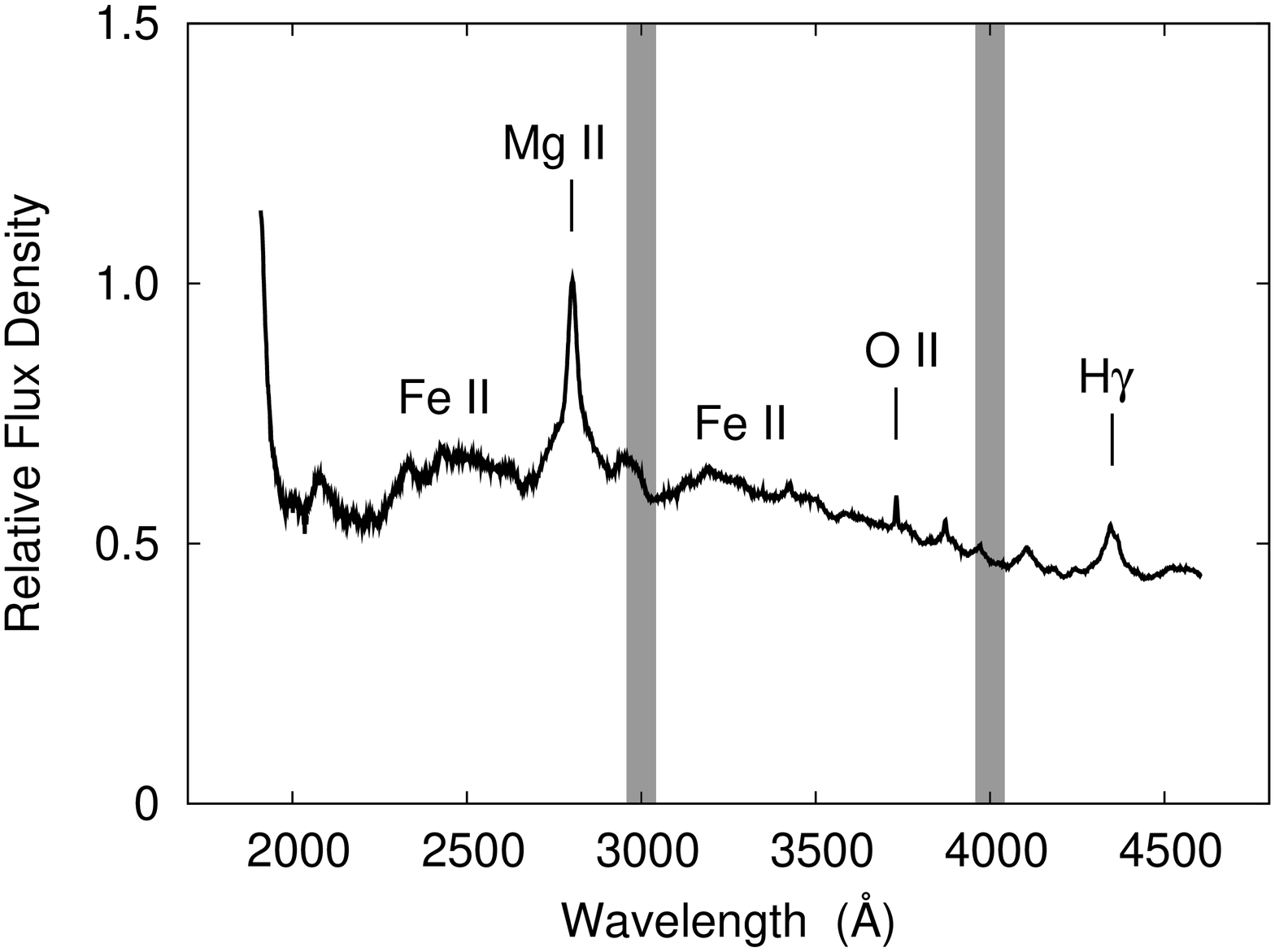}}
\end{tabular}
\end{center}
\caption{
\label{spec_fluxratio}
An example of the spectrum of an SDSS QSO at the rest frame.
Two gray bars indicate the wavebands of stacked-flux 
at $\lambda=3000\pm50\AA$~and $4000\pm50\AA$.\\
}
\end{figure}

Our QSOs were selected from the SDSS DR7 Quasar Catalog \citep{sch10}.
We limited our sample to $i$-band magnitudes brighter than 
 the PSF magnitude of 19.1 \footnote[1]{http://www.sdss2.org/dr7/ PSF mag. $i<19.1$ for $z<2.3$}.
We downloaded the spectra\footnote[2]{
The spectra were downloaded using the interface of the Tenth SDSS Data Release. 
We note that the spectra are actually from DR7, 
 which mainly includes QSOs at $z<2.3$.
For DR10, the QSOs are at $z>2.2$.
}
 of these QSOs and
 defined the QSO color with the flux ratio of the two continuum bands
 at the rest frame 
 $\lambda=4000\pm50$\AA~to $\lambda=3000\pm50$\AA~(see Figure~\ref{spec_fluxratio}).
The flux ratio $r$ is defined as,

\begin{equation}
\label{eq:fluxratio}
r = \frac{\displaystyle {\bar{f}_{4000}}}
  {\displaystyle  {\bar{f}_{3000}}} \\
= \frac{\displaystyle \int^{4050\mathrm\AA}_{3950\mathrm\AA} {f_{\lambda}d\lambda}}
  {\displaystyle  \int^{3050\mathrm\AA}_{2950\mathrm\AA} {f_{\lambda}d\lambda}},
\end{equation}
where
 {$\bar f_\lambda$} is the mean flux density within a range of wavelengths, 
 $f_\lambda$ is the flux density per unit wavelength, 
 and $\lambda$ is in units of \AA.
These two bands were selected to avoid strong narrow lines (e.g., \ion{Mg}{2}, \ion{O}{2}, and H$\gamma$)
 and broad emission lines (e.g., \ion{Fe}{2}). 
Because of the  wavelength range of SDSS spectra, 
 only QSOs with redshifts from 0.3 to 1.2 could be selected.

\begin{figure}
\begin{center}
\begin{tabular}{cc}
{\includegraphics[height=8.5cm,angle=-90]{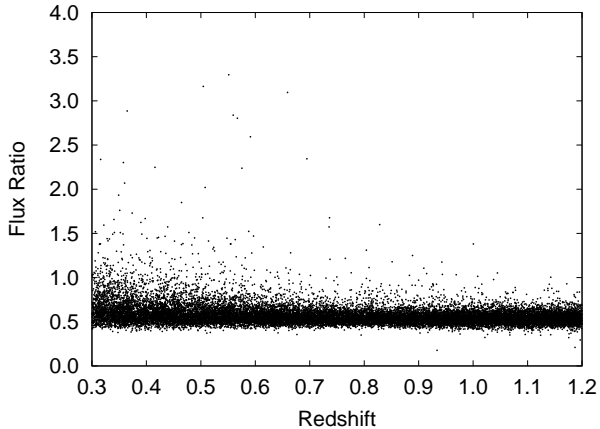}}
\end{tabular}
\end{center}
\caption{
\label{fluxratio_z}
Flux ratio of the QSO samples at different redshifts. 
\\
}
\end{figure}

\begin{figure}
\begin{center}
\begin{tabular}{cc}
{\includegraphics[height=8.5cm,angle=-90]{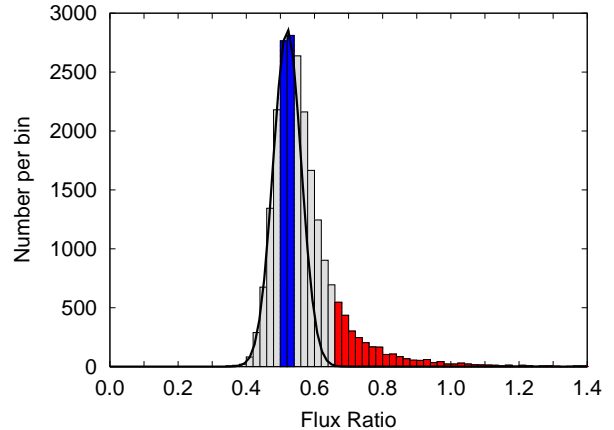}}
\end{tabular}
\end{center}
\caption{
\label{fluxratio_number}
Number distributions of the QSO samples versus the continuum flux ratios
 of 4000\AA~to 3000\AA~within the FIRST sky coverage.
The number bin width is 0.02.
The black curve is a Gaussian fitting with peak center at flux ratio $r_{\rm peak} = 0.52$,
 and $\sigma\approx   0.042$.
We select QSOs with flux ratio 
 $r_{\rm peak}\pm\frac{1}{2}\sigma$, i.e., 
 $0.50 \leqslant r \leqslant 0.54$
as tQSOs (blue color),
 and
QSOs with flux ratio $r\geqslant 0.66$  as rQSOs (red color).
The other QSOs are shown in gray color.
\\
}
\end{figure}

Figure~\ref{fluxratio_z} shows the flux ratios at different redshifts.
We found that the mode values of flux ratios at different redshifts
 are all located within 0.5 -- 0.6 bin and were independent of redshift.
Notably, the flux ratios at low redshifts 
 more scattered than those at high redshifts.
The number distribution of the flux ratios look like a Gaussian distribution
 with an extended right tail (Figure~\ref{fluxratio_number}).
Therefore, we fitted the left wing of the distribution with a Gaussian function
 and folded the result to the right wing.
The distribution peaked at $r_{\rm peak} = 0.52$ with $\sigma\approx   0.042$.
The 3$\sigma$ at the right wing was at $r\approx 0.65$.
Therefore, for comparison, we defined our QSO samples with $r \geqslant 0.66$ 
 as red QSOs (hereafter, rQSOs),
and the QSOs with $0.50\leqslant r \leqslant 0.54$ ($r_{\rm peak}\pm\frac{1}{2}\sigma$) 
 as typical QSOs (hereafter, tQSOs).
Notably, the tQSOs only represented
 the sources with the most likely color in the SDSS QSO samples.
Section~\ref{discussion} offers further discussion
 regarding the distribution of the QSO samples.
We noted that the color ratio difference between
 the the rQSOs and the tQSOs 
 roughly corresponded to absorption of 0.26 magnitude at 4000\AA~when
 the colors were caused by dust absorption,
 assuming that the SMC dust extinction law \citep{pei92} was in effect.
Notably, the effective E(B--V) of 0.26 was similar to
  the definition of red QSOs in \citet{lac13,lac15}.

\begin{figure}
\begin{center}
\begin{tabular}{cc}
{\includegraphics[height=8.5cm,angle=-90]{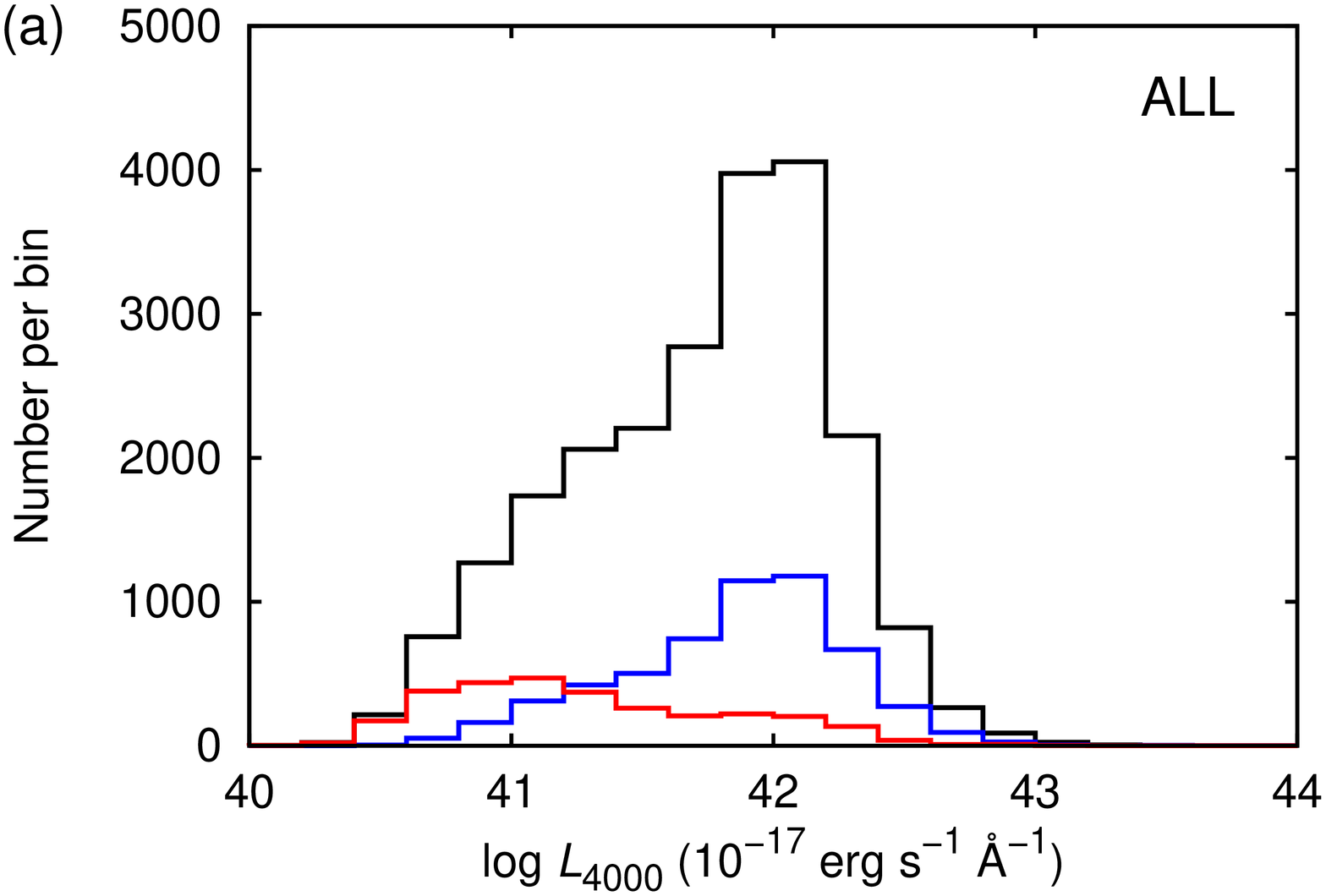}}\\
{\includegraphics[height=8.5cm,angle=-90]{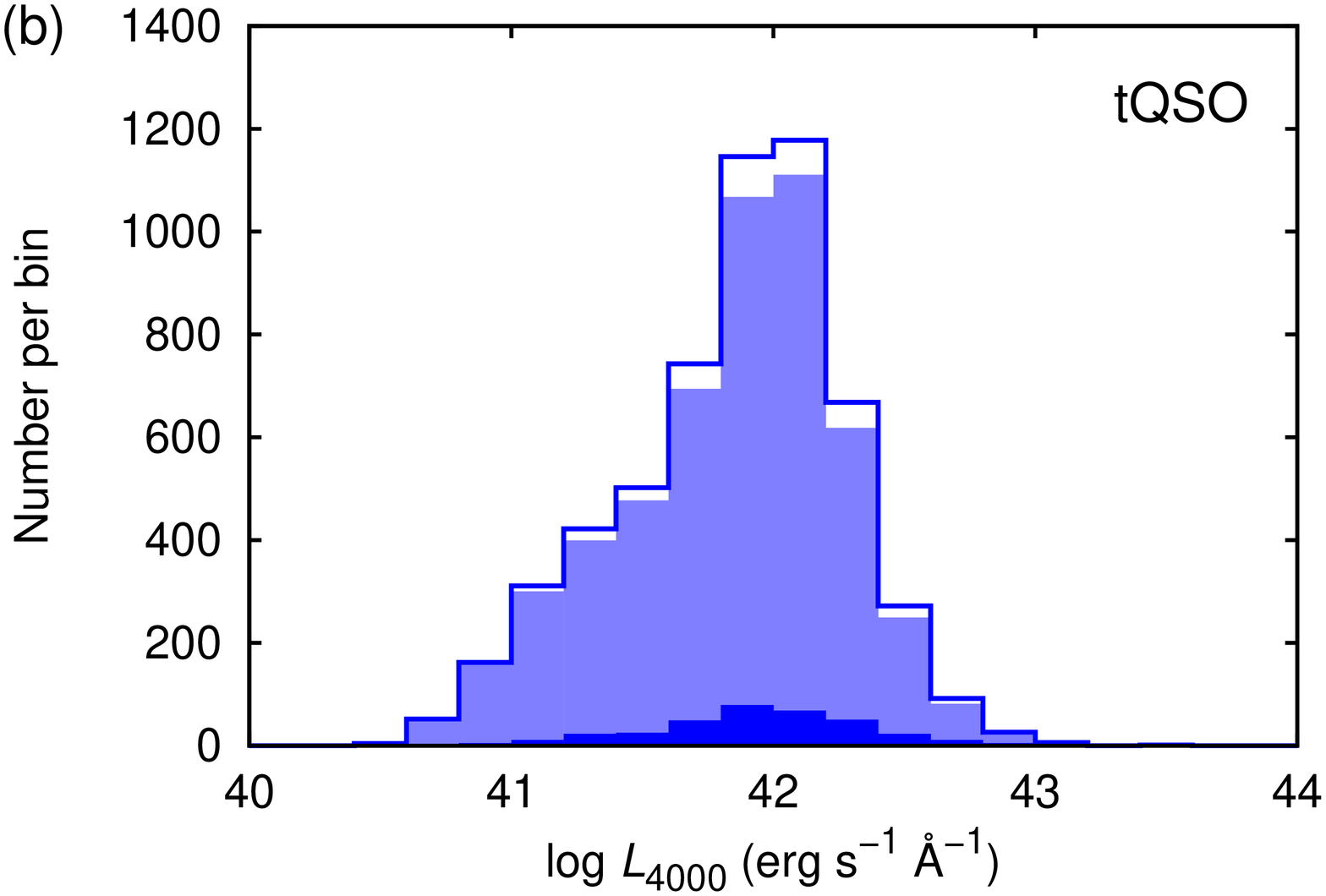}}\\
{\includegraphics[height=8.5cm,angle=-90]{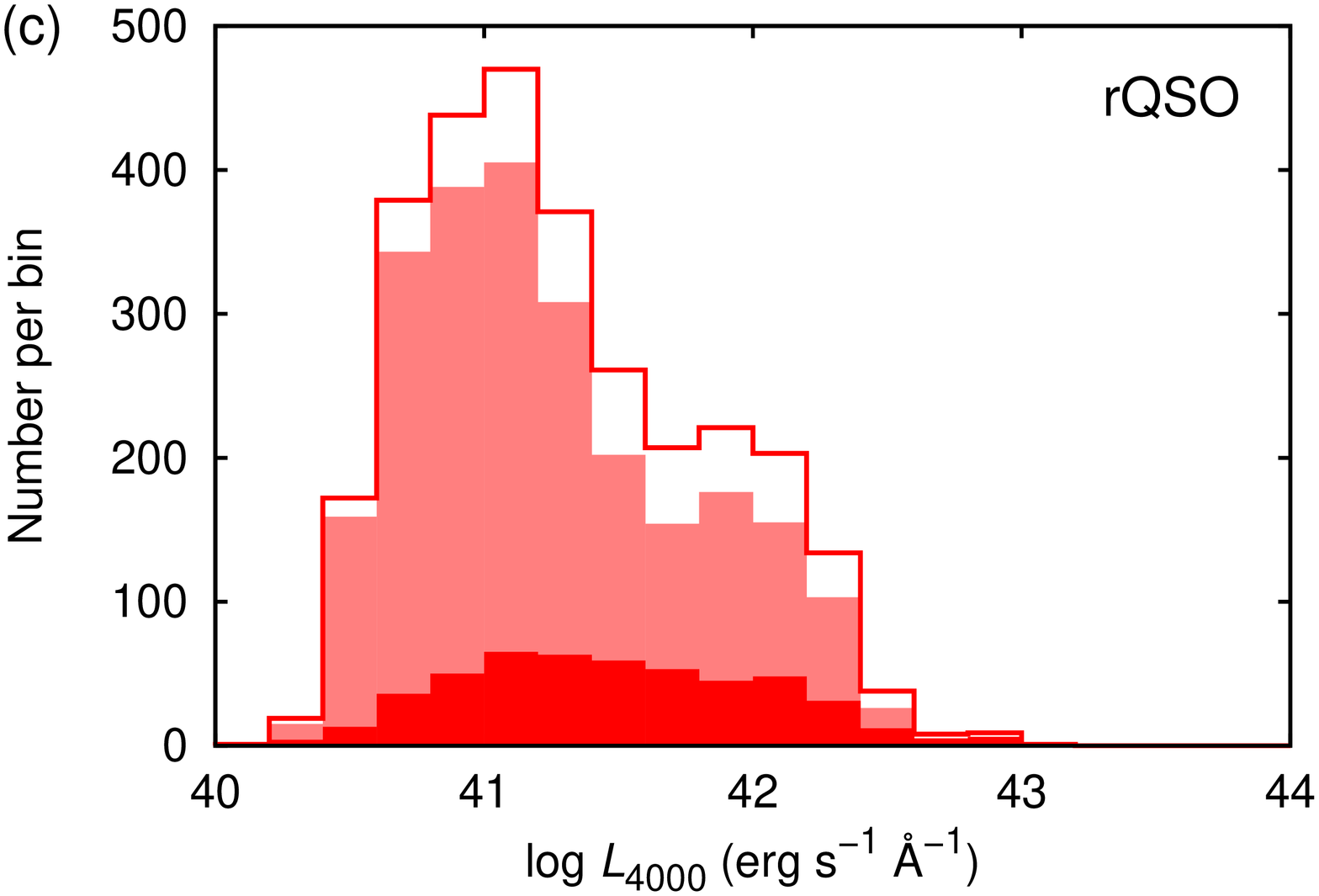}}
\end{tabular}
\end{center}
\caption{
\label{lum_number_rlqrqq}
Number distributions of different types of QSOs at different luminosities.
{\bf (a)} 
Comparison between the tQSOs and the rQSOs.
The black solid line represents the whole QSO sample,
 the blue solid line represents the tQSOs, and
 the red solid line represents the rQSOs.
{\bf (b)} 
Comparison between the radio-quiet and the radio-loud sources within the tQSOs.
The blue solid line represents the tQSOs,
 the filled light-blue histogram represents the tRQQs, and
 the filled blue histogram represents the tRLQs.
{\bf (c)} 
Comparison between the radio-quiet and the radio-loud sources within the rQSOs.
The red solid line represents the rQSOs,
 the filled light-red histogram represents the rRQQs, and
 the filled red histogram represents the rRLQs.
\\}
\end{figure}

\begin{figure*}
\begin{center}
\begin{tabular}{cc}
{\includegraphics[height=8.5cm,angle=-90]{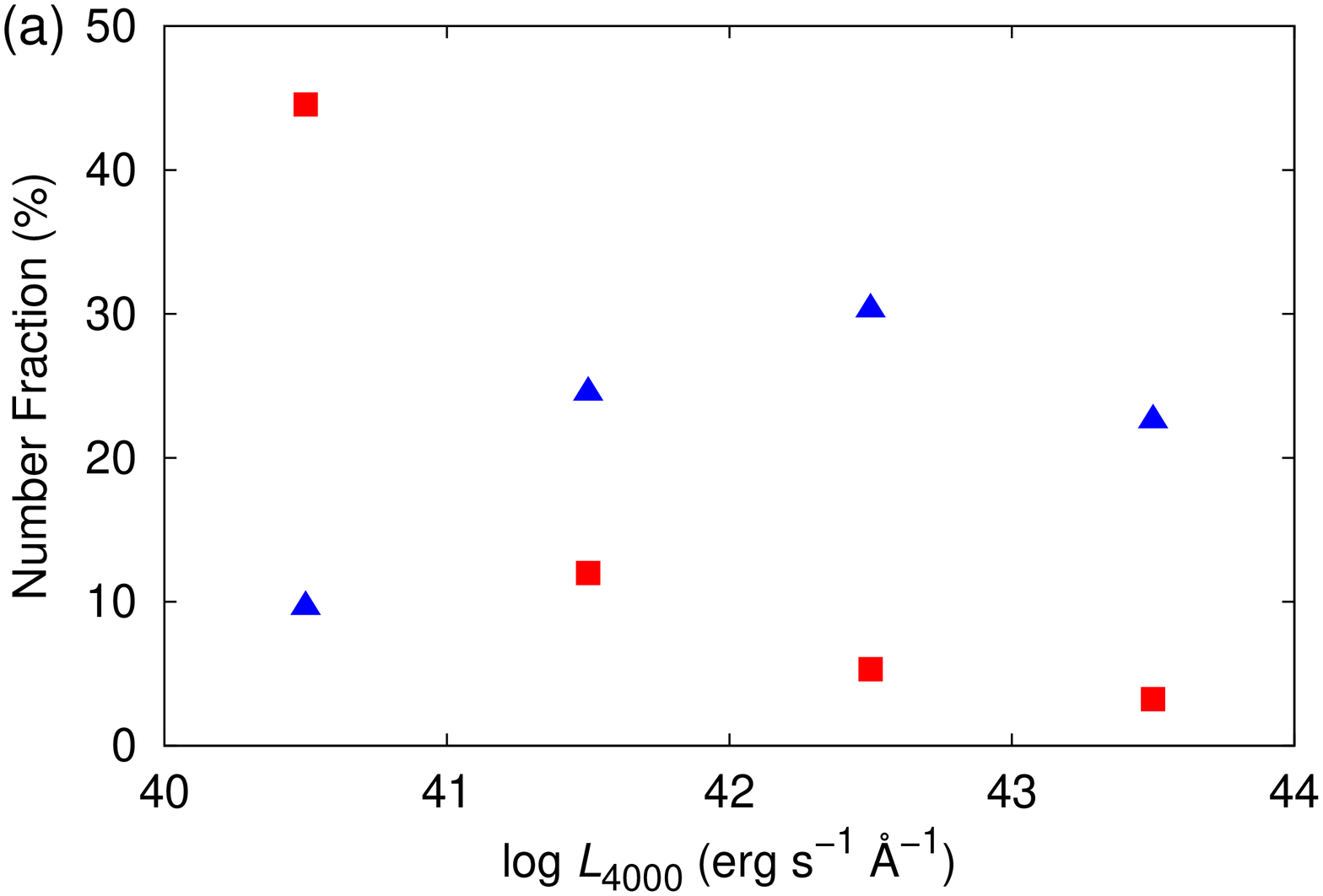}}
{\includegraphics[height=8.5cm,angle=-90]{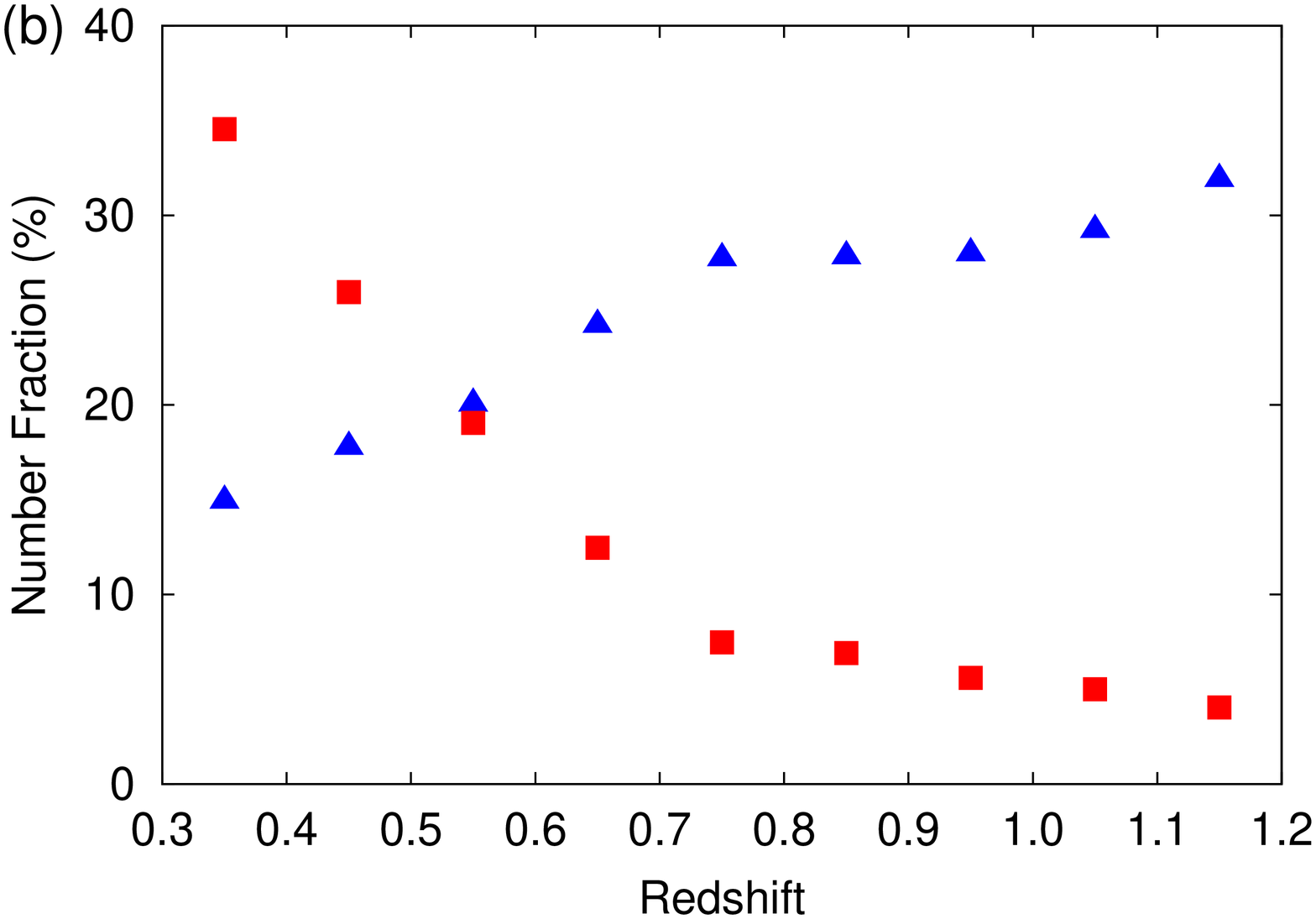}}
\end{tabular}
\end{center}
\caption{
\label{red_fraction}
{\bf (a)} Number fractions of
 the tQSOs (filled blue triangles) and
 the rQSOs (filled red squares) 
 with different luminosities.
{\bf (b)} Number fractions of
 the tQSOs (filled blue triangles) and
 the rQSOs (filled red squares)
 at different redshifts.
\\}
\end{figure*}

\begin{figure*}
\begin{center}
\begin{tabular}{cc}
{\includegraphics[height=8.5cm,angle=-90]{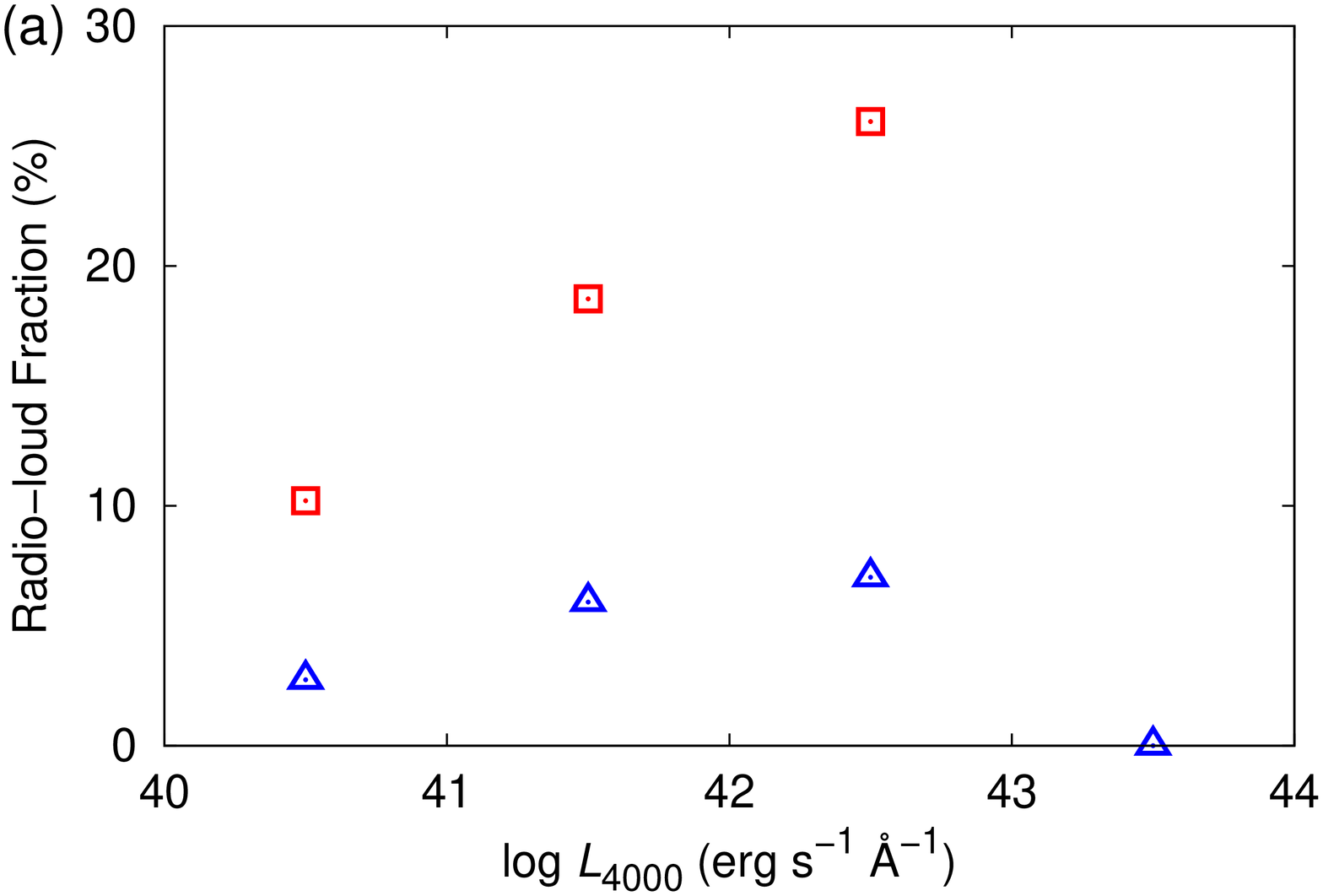}}
{\includegraphics[height=8.5cm,angle=-90]{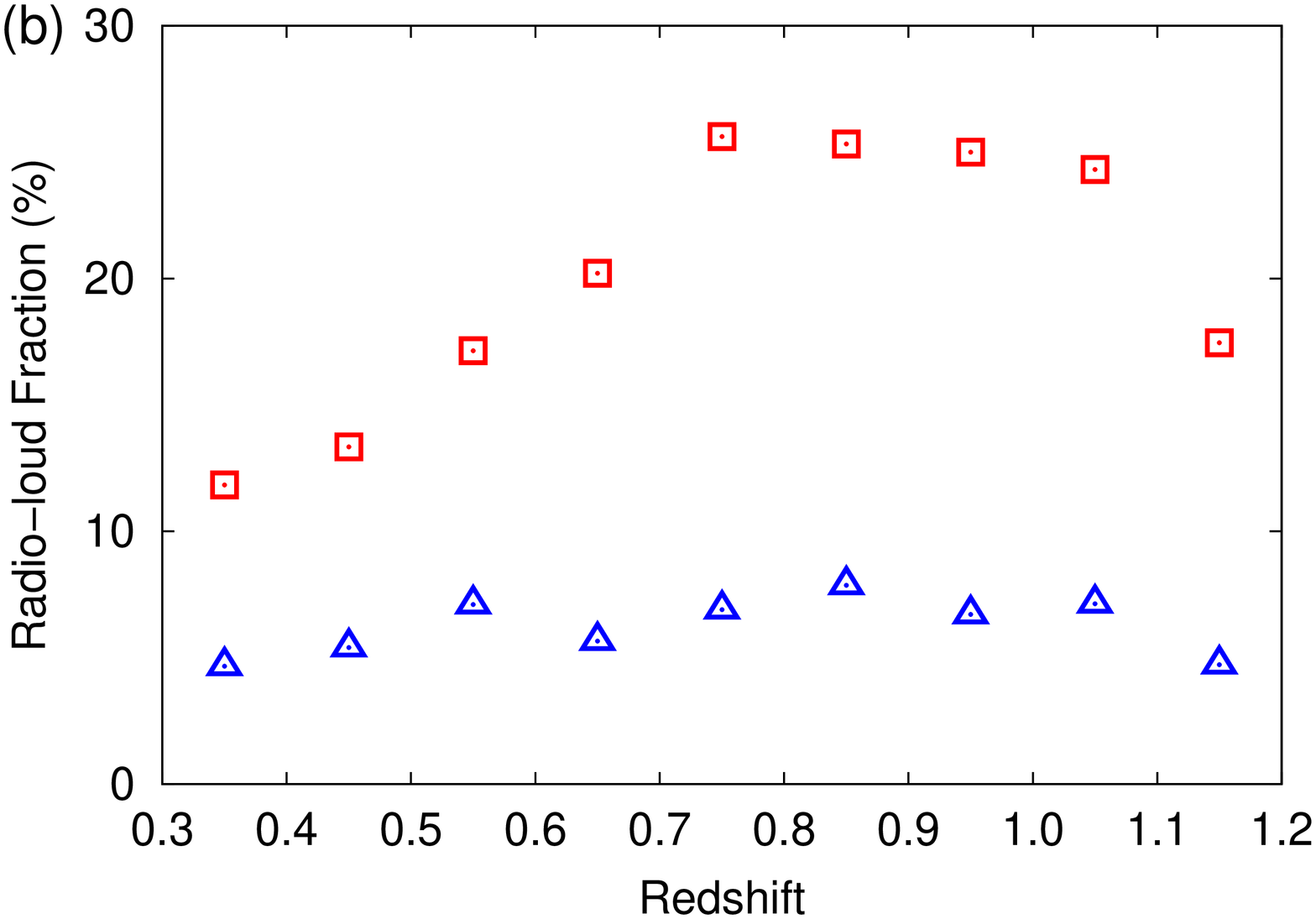}}
\end{tabular}
\end{center}
\caption{
\label{radioloud_fraction}
{\bf (a)}
Radio-loud fractions of
 the tQSOs (open blue triangles) and
 the rQSOs (open red squares) 
  with different luminosities.
The data points at log$(L_{4000})>43$ can be ignored
 due to meaningless statistics (small sample size).
{\bf (b)}
Radio-loud fractions of
 the tQSOs (open blue triangles) and
 the rQSOs (open red squares) 
 with different redshifts.
The symbols are the same as those in the left panel.
\\}
\end{figure*}

\begin{table*}
\begin{center}
\caption{Number and number fractions of tQSOs and rQSOs at different luminosities}
\label{tab_fraction_lum}
\begin{tabular}{cccccc}
\hline
\hline
Luminosity & $n_{\rm QSO}$ & $n_{\rm tQSO}$ & $\xi(\rm tQSO)$ & $n_{\rm rQSO}$ & $\xi(\rm rQSO)$  \\
\hline
40 $\leqslant {\rm log}L_{4000} <$ 41  &    \enspace2265  &    \enspace218  & \enspace9.62\% &           1009 & 44.55\% \\
41 $\leqslant {\rm log}L_{4000} <$ 42  &           12749  &           3124  &        24.50\% &           1530 & 12.00\% \\
42 $\leqslant {\rm log}L_{4000} <$ 43  &    \enspace7384  &           2236  &        30.28\% &    \enspace392 & \enspace5.31\% \\
43 $\leqslant {\rm log}L_{4000} <$ 44  & \hskip 1.5em 31  & \hskip 1.5em 7  &        22.58\% & \hskip 1.5em 1 & \enspace3.23\% \\
\hline
40 $\leqslant {\rm log}L_{4000} <$ 44  &   22429  &    5585  &   24.90\% & 2932 & 13.07\% \\
\hline
\multicolumn{6}{l}{1. $n_{\rm QSO}$ is the number of the whole QSO sample.}\\
\multicolumn{6}{l}{2. $n_{\rm tQSO}$ is the number of the typical QSOs.
$\xi(\rm tQSO) = n_{\rm tQSO}/n_{\rm QSO}$.}\\
\multicolumn{6}{l}{3. $n_{\rm rQSO}$ is the number of the red QSOs.
$\xi(\rm rQSO) = n_{\rm rQSO}/n_{\rm QSO}$.}\\
\end{tabular}
\end{center}
\end{table*}

\begin{table*}
\begin{center}
\caption{Number and number fractions of tQSOs and rQSOs at different redshifts}
\label{tab_fraction_z}
\begin{tabular}{cccccc}
\hline
\hline
Redshift & $n_{\rm QSO}$ & $n_{\rm tQSO}$ & $\xi(\rm tQSO)$ & $n_{\rm rQSO}$ & $\xi(\rm rQSO)$  \\
\hline
0.3 $\leqslant z <$ 0.4  & \enspace2446  & \enspace365  &   14.92\% & \enspace845  &   34.55\% \\
0.4 $\leqslant z <$ 0.5  & \enspace2397  & \enspace426  &   17.77\% & \enspace622  &   25.95\% \\
0.5 $\leqslant z <$ 0.6  & \enspace2389  & \enspace479  &   20.05\% & \enspace455  &   19.05\% \\
0.6 $\leqslant z <$ 0.7  & \enspace2264  & \enspace548  &   24.20\% & \enspace282  &   12.46\% \\
0.7 $\leqslant z <$ 0.8  & \enspace2143  & \enspace594  &   27.72\% & \enspace160  & \enspace7.47\% \\
0.8 $\leqslant z <$ 0.9  & \enspace2286  & \enspace636  &   27.82\% & \enspace158  & \enspace6.91\% \\
0.9 $\leqslant z <$ 1.0  & \enspace2502  & \enspace700  &   27.98\% & \enspace140  & \enspace5.60\% \\
1.0 $\leqslant z <$ 1.1  & \enspace2880  & \enspace841  &   29.20\% & \enspace144  & \enspace5.00\% \\
1.1 $\leqslant z <$ 1.2  & \enspace3122  & \enspace996  &   31.90\% & \enspace126  & \enspace4.04\% \\
\hline
0.3 $\leqslant z <$ 1.2  &   22429  &    5585  &   24.90\% &     2932  &   13.07\% \\
\hline
\multicolumn{6}{l}{1. The symbols are the same with those in Table~\ref{tab_fraction_lum}.}\\
\end{tabular}
\end{center}
\end{table*}

\begin{table*}
\begin{center}
\caption{Number and number fractions of tRLQs and rRLQs at different luminosities}
\label{tab_fraction_lum_rlq}
\begin{tabular}{cccccc}
\hline
\hline
Luminosity & $n_{\rm RLQ}$ & $n_{\rm tRLQ}$ & $\xi_{\rm RLQ}(\rm tRLQ)$ & $n_{\rm rRLQ}$ & $\xi_{\rm RLQ}(\rm rRLQ)$ \\
\hline
40 $\leqslant {\rm log}(L_{4000}) <$ 41  &     159  & \quad6  & \enspace3.77\% &      103    &   64.78\% \\
41 $\leqslant {\rm log}(L_{4000}) <$ 42  &     992  &     187 &        18.85\% &      285    &   28.73\% \\
42 $\leqslant {\rm log}(L_{4000}) <$ 43  &     674  &     157 &        23.29\% &      102    &   15.13\% \\
43 $\leqslant {\rm log}(L_{4000}) <$ 44  &  \quad4  &  \quad0 & \enspace0.00\% & \quad1    &   25.00\% \\
\hline
40 $\leqslant {\rm log}(L_{4000}) <$ 44  &    1829  &     350 &   19.14\% &      491  &   26.85\% \\
\hline
\multicolumn{6}{l}{1. $n_{\rm RLQ}$ is the number of the radio-loud QSOs.}\\
\multicolumn{6}{l}{2. $n_{\rm tRLQ}$ is the number of the radio-loud typical QSOs.}
$\xi_{\rm RLQ}(\rm tRLQ) = n_{\rm tRLQ}/n_{\rm RLQ}$.\\
\multicolumn{6}{l}{3. $n_{\rm rRLQ}$ is the number of the radio-loud red QSOs.}
$\xi_{\rm RLQ}(\rm rRLQ) = n_{\rm rRLQ}/n_{\rm RLQ}$.\\
\end{tabular}
\end{center}
\end{table*}

\begin{table*}
\begin{center}
\caption{Number and number fractions of tRLQs and rRLQs at different redshifts}
\label{tab_fraction_z_rlq}
\begin{tabular}{cccccc}
\hline
\hline
Redshift & $n_{\rm RLQ}$ & $n_{\rm tRLQ}$ & $\xi_{\rm RLQ}(\rm tRLQ)$ & $n_{\rm rRLQ}$ & $\xi_{\rm RLQ}(\rm rRLQ)$ \\
\hline
0.3 $\leqslant z <$ 0.4  & \enspace191  & \enspace17  & \enspace8.90\% &      100  &   52.36\% \\
0.4 $\leqslant z <$ 0.5  & \enspace183  & \enspace23  &   12.57\% & \enspace83  &   45.36\% \\
0.5 $\leqslant z <$ 0.6  & \enspace211  & \enspace34  &   16.11\% & \enspace78  &   36.97\% \\
0.6 $\leqslant z <$ 0.7  & \enspace175  & \enspace31  &   17.71\% & \enspace57  &   32.57\% \\
0.7 $\leqslant z <$ 0.8  & \enspace197  & \enspace41  &   20.81\% & \enspace41  &   20.81\% \\
0.8 $\leqslant z <$ 0.9  & \enspace192  & \enspace50  &   26.04\% & \enspace40  &   20.83\% \\
0.9 $\leqslant z <$ 1.0  & \enspace203  & \enspace47  &   23.15\% & \enspace35  &   17.24\% \\
1.0 $\leqslant z <$ 1.1  & \enspace251  & \enspace60  &   23.90\% & \enspace35  &   13.94\% \\
1.1 $\leqslant z <$ 1.2  & \enspace226  & \enspace47  &   20.80\% & \enspace22  & \enspace9.73\% \\
\hline
0.3 $\leqslant z <$ 1.2  &    1829  &     350  &   19.14\% &      491  &   26.85\% \\
\hline
\multicolumn{6}{l}{1. The symbols are the same with those in Table~\ref{tab_fraction_lum_rlq}.}\\
\end{tabular}
\end{center}
\end{table*}

\begin{table}
\begin{center}
\caption{Radio-loud fractions of tQSOs and rQSOs at different luminosities}
\label{tab_rlq_lum}
\begin{tabular}{cccc}
\hline
\hline
Luminosity & $\xi(\rm RLQ)$ & $\xi_{\rm tQSO}(\rm tRLQ)$ & $\xi_{\rm rQSO}(\rm rRLQ)$ \\
\hline
40 $\leqslant {\rm log}(L_{4000}) <$ 41 &    \enspace7.02\% &     2.75\% &   \enspace10.21\% \\
41 $\leqslant {\rm log}(L_{4000}) <$ 42 &    \enspace7.78\% &     5.99\% &   \enspace18.63\% \\
42 $\leqslant {\rm log}(L_{4000}) <$ 43 &    \enspace9.13\% &     7.02\% &   \enspace26.02\% \\
43 $\leqslant {\rm log}(L_{4000}) <$ 44 &    12.90\% &     0.00\% &   100.00\% \\
\hline
40 $\leqslant {\rm log}(L_{4000}) <$ 44 &    \enspace8.15\% &     6.27\% &   \enspace16.75\% \\
\hline
\multicolumn{4}{l}{1. $\xi(\rm RLQ) = n_{\rm RLQ}/n_{\rm QSO}$.}\\
\multicolumn{4}{l}{2. $\xi_{\rm tQSO}(\rm tRLQ) = n_{\rm tRLQ}/n_{\rm tQSO}$.}\\
\multicolumn{4}{l}{3. $\xi_{\rm rQSO}(\rm rRLQ) = n_{\rm rRLQ}/n_{\rm rQSO}$.}\\
\end{tabular}
\end{center}
\end{table}

\begin{table}
\begin{center}
\caption{Radio-loud fraction of tQSOs and rQSOs at different redshifts}
\label{tab_rlq_z}
\begin{tabular}{cccc}
\hline
\hline
Redshift & $\xi(\rm RLQ)$ & $\xi_{\rm tQSO}(\rm tRLQ)$ & $\xi_{\rm rQSO}(\rm rRLQ)$ \\
\hline
0.3 $\leqslant z <$ 0.4 &     7.81\% &     4.66\% &    11.83\% \\
0.4 $\leqslant z <$ 0.5 &     7.63\% &     5.40\% &    13.34\% \\
0.5 $\leqslant z <$ 0.6 &     8.83\% &     7.10\% &    17.14\% \\
0.6 $\leqslant z <$ 0.7 &     7.73\% &     5.66\% &    20.21\% \\
0.7 $\leqslant z <$ 0.8 &     9.19\% &     6.90\% &    25.62\% \\
0.8 $\leqslant z <$ 0.9 &     8.40\% &     7.86\% &    25.32\% \\
0.9 $\leqslant z <$ 1.0 &     8.11\% &     6.71\% &    25.00\% \\
1.0 $\leqslant z <$ 1.1 &     8.72\% &     7.13\% &    24.31\% \\
1.1 $\leqslant z <$ 1.2 &     7.24\% &     4.72\% &    17.46\% \\
\hline
0.3 $\leqslant z <$ 1.2 &     8.15\% &     6.27\% &    16.75\% \\
\hline
\multicolumn{4}{l}{1. The symbols are the same with those in Table~\ref{tab_rlq_lum}.}\\
\end{tabular}
\end{center}
\end{table}

\begin{figure*}
\begin{center}
\begin{tabular}{cc}
{\includegraphics[height=17cm,angle=-90]{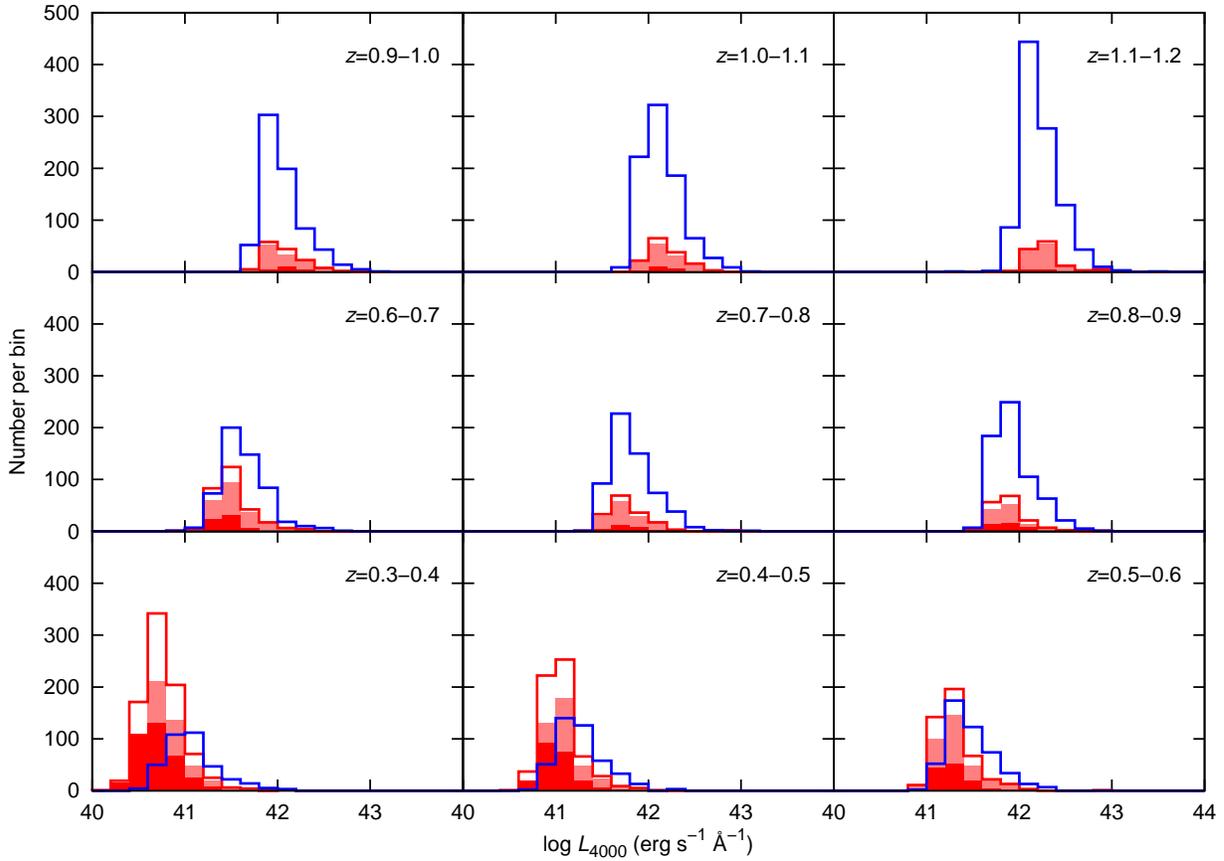}}
\end{tabular}
\end{center}
\caption{
\label{lum_number_red_norm_3x3}
Number distributions of different types of QSOs versus different luminosities at different redshifts.
The blue solid line represents the tQSOs with flux ratios of 
  $0.50 \leqslant r \leqslant 0.54$.
The red solid line represents the rQSOs with flux ratios  $\geqslant 0.66$.
The filled light-red histogram represents the rQSOs with flux ratios of
  $0.66 \leqslant r< 0.80$.
The filled red histogram represents the rQSOs with flux ratios $\geqslant 0.80$.
See more information of different types of QSOs in Table~\ref{stddev_qso}.
\\}
\end{figure*}

\begin{figure}
\begin{center}
\begin{tabular}{cc}
{\includegraphics[height=8.5cm,angle=-90]{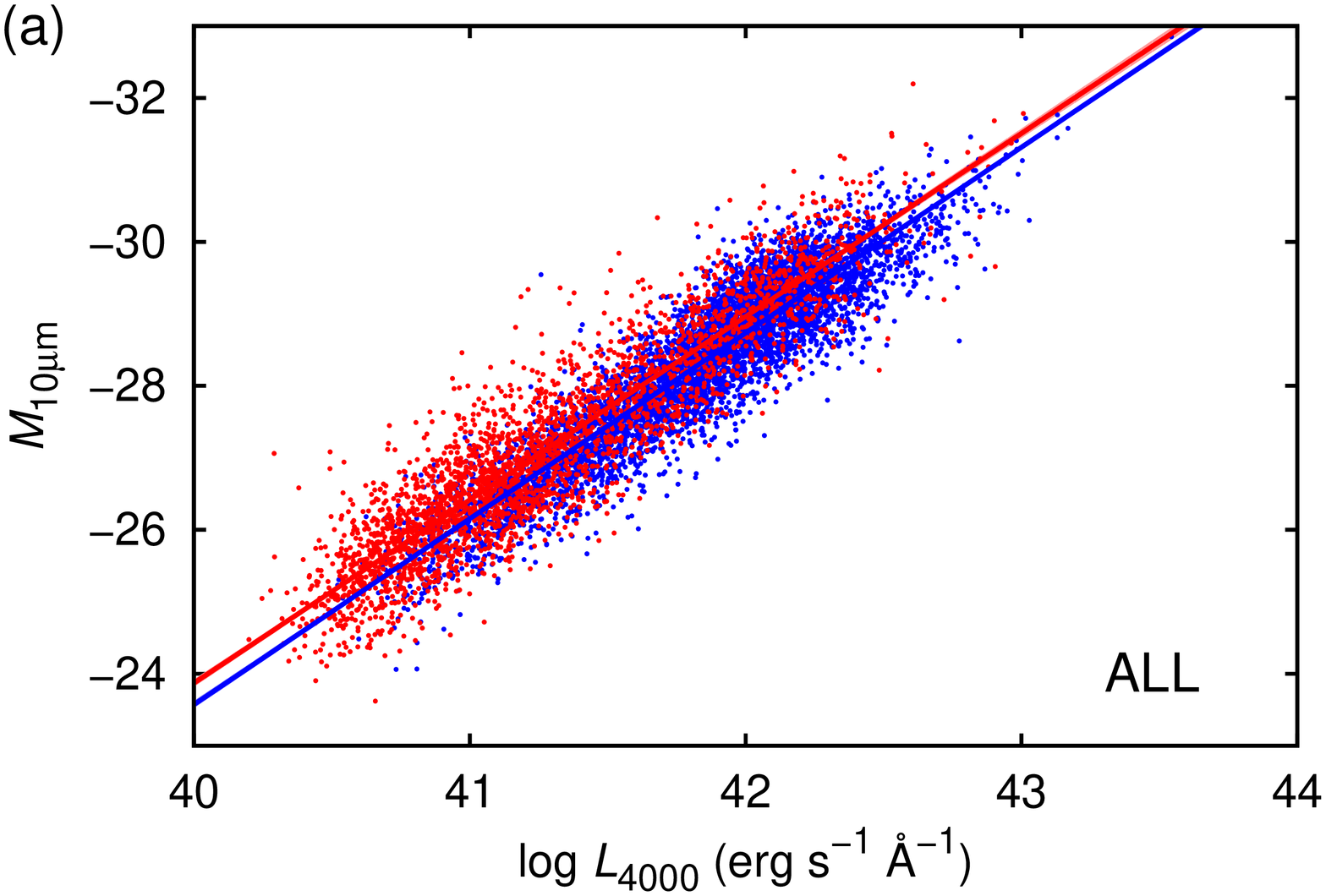}}\\
{\includegraphics[height=8.5cm,angle=-90]{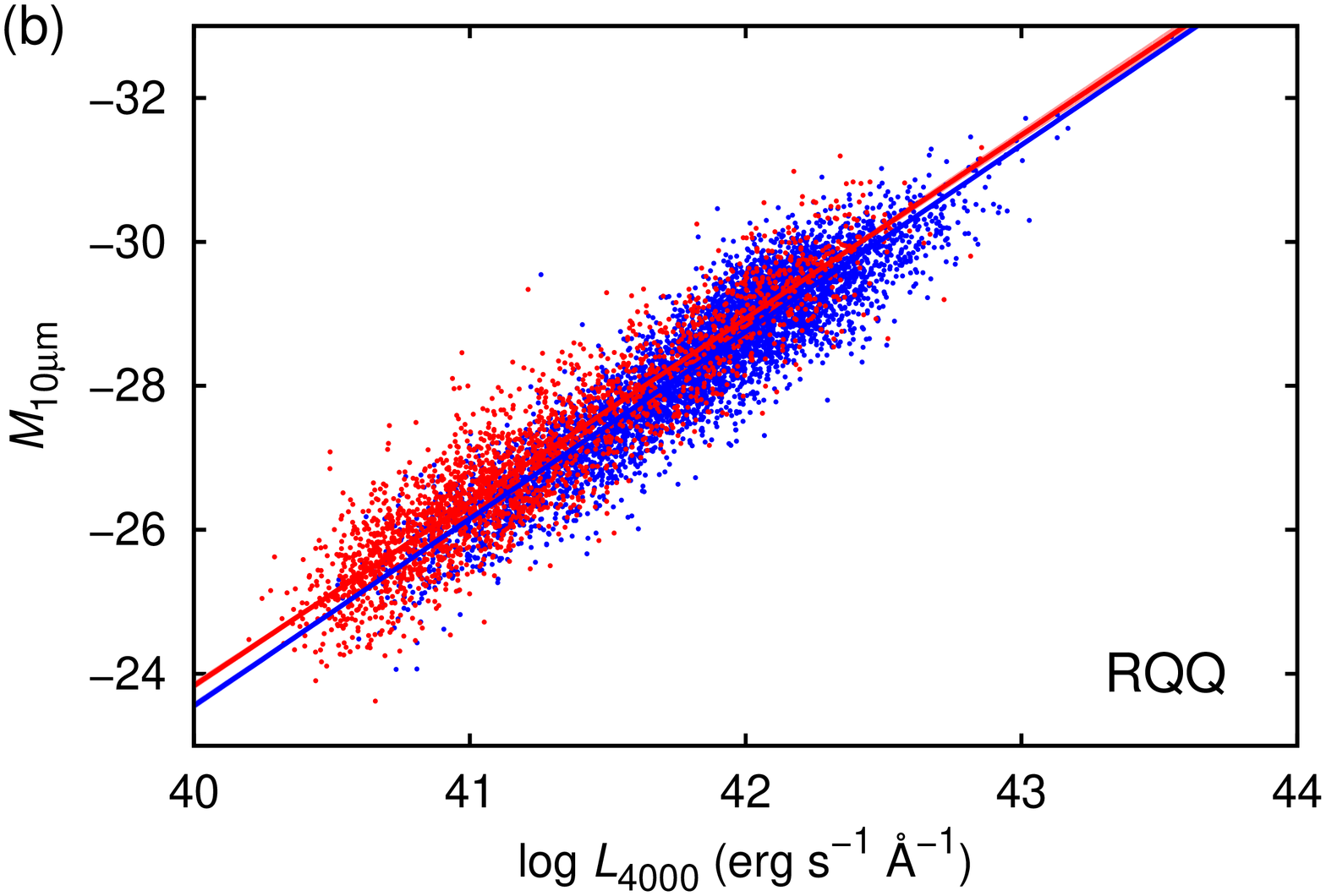}}\\
{\includegraphics[height=8.5cm,angle=-90]{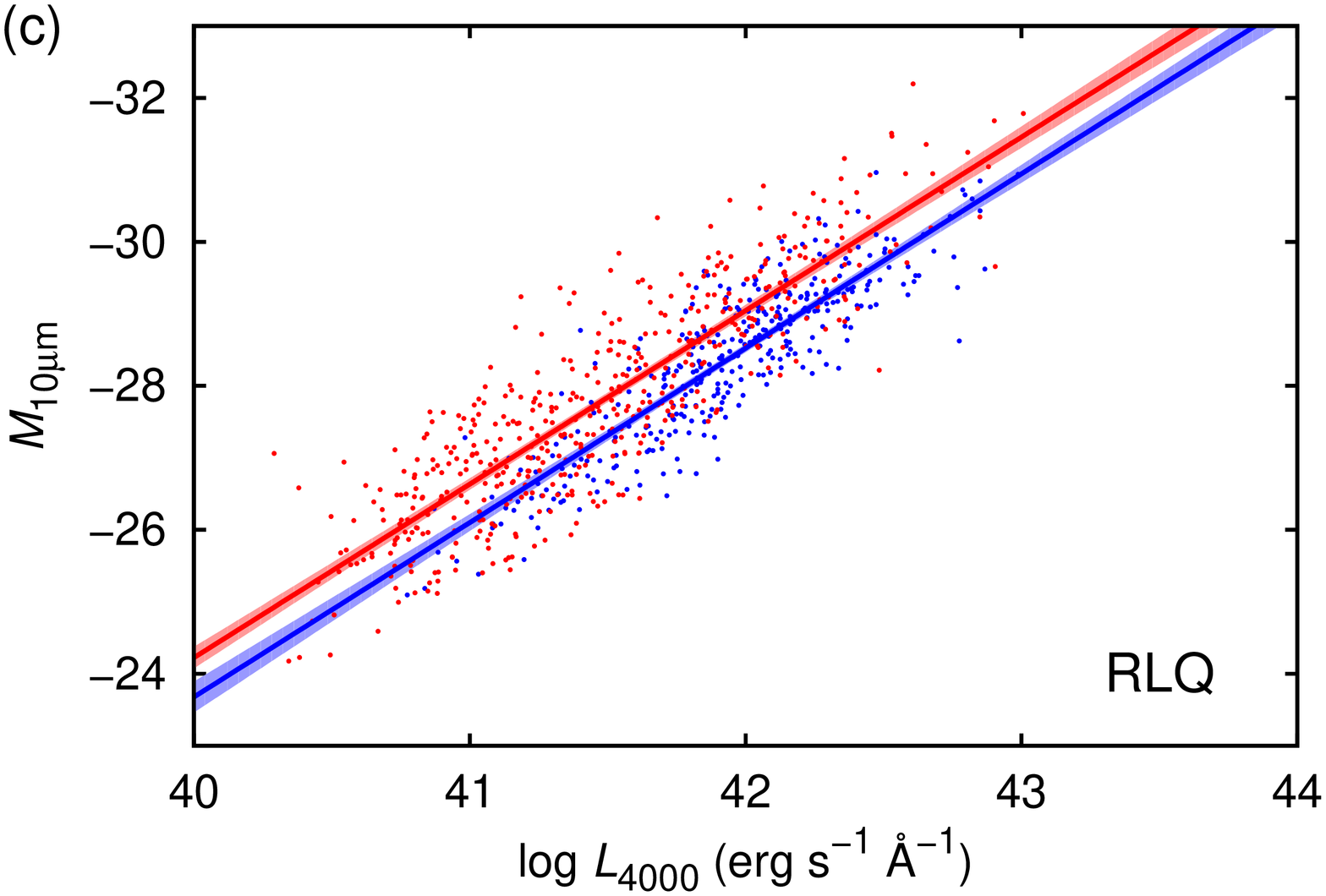}}\\
\end{tabular}
\end{center}
\caption{
\label{lum_10micron}
Absolute magnitude of 10\micron~vs. log($L_{4000}$) 
for {\bf (a)} the whole QSO sample, {\bf (b)} the RQQs, and {\bf (c)} the RLQs.
Blue dots are the tQSOs, and red dots are the rQSOs.
Blue line and red line are
 linear $\chi^2$ fitting of the tQSOs and the rQSOs,
 respectively.
Blue shadow and red shadow in each figure
 are 95\% confidence intervals of the tQSOs and the rQSOs,
 respectively.
Note that the confidence intervals are too small to be easily visible for (a) and (b),
 but is still visible for (c).
\\}
\end{figure}

\subsection{\rm Radio Data and Definition of Radio-loud QSOs and Radio-quiet QSOs} 
We looked for the radio counterparts of the QSOs 
in the sky coverage of the VLA FIRST radio survey \citep{bec97}
within a 2$\arcsec$ radius,
 which can match 98\% of sources with real association (e.g., \citealt{mcm02}.)
For the QSOs with radio detection, 
 we used their radio-to-optical ratios to classify a group of radio-loud QSOs (RLQs) \citep{kel89,ive02}.
The radio-to-optical ratio $R_{\rm g}$ is defined as
\begin{equation}
R_{\rm g} = {\rm log}(\frac{f_{\rm radio}}{f_{\rm optical}}) = 0.4(g-t),
\end{equation}
where $f_{\rm radio}$ is the radio flux,
 $f_{\rm optical}$ is the optical flux,
 $g$ is the SDSS $g$-band magnitude,
 and $t$ is the radio magnitude defined as
\begin{equation}
t=-2.5{\rm\ log}(\frac{f_{\rm FIRST}}{3631~{\rm Jy}}),
\end{equation}
where $f_{\rm FIRST}$ is the integrated flux density of FIRST
\citep{ive02,oke83}.
We defined QSOs with $R_{\rm g}>1$ as RLQs \citep{kel89,ive02}.
The QSOs without radio detection
 and those with radio detection
 but with $R_{\rm g}\leqslant 1$ 
 were both classified as radio-quiet QSOs (RQQs).

\subsection{\rm Infrared Data} 
In order to investigate the dust properties at the infrared band,
 we searched for the infrared counterpart of the QSOs 
 within the FIRST sky coverage using the data of WISE,
which performed an all sky survey at four different wavebands,
 W1 (3.4~\micron), W2 (4.6~\micron),
 W3 (12~\micron), and W4 (22~\micron),
 where W1 and W2 are narrow bands,
 and W3 and W4 are broad bands
 \citep{wri10}.
The infrared counterparts were identified
 by matching the All-Sky WISE Source Catalog objects within a 2$\farcs$75 radius,
 which is the pixel scale of the WISE detectors.
We found a counterpart in WISE within a 2$\farcs$75 radius
 search for all SDSS QSOs, except for 22 sources.
For the following analysis, we considered only these WISE-detected sources.

\section{Results}
\subsection{\rm Distributions of the rQSOs}
Our sample included 22,429 QSOs, of which 5,585 were tQSOs
  with $0.50 \leqslant r \leqslant 0.54$ ($\approx 25\%$ of the total QSOs)
 and 2,932 were rQSOs 
  with $r\geqslant 0.66$ ($\approx 13\%$ of the total QSOs).
In order to compare the luminosities of different types of QSOs,
 we chose the luminosity at the rest frame 4000\AA~to
 represent the luminosity of each source.
The luminosity per unit wavelength can be derived from the flux density per unit wavelength  \citep{hog99},
\begin{equation}
\label{eq:flambda}
f_{\lambda} = \frac{1}{1+z}\frac{L_{\lambda/(1+z)}}{L_{\lambda}}\frac{L_{\lambda}}{4\pi~D_L^{2}},
\end{equation}
  where 
   $z$ is the redshift, 
 $L_{\lambda/(1+z)}$ is the luminosity per unit wavelength at rest frame,
 $L_{\lambda}$ is the luminosity per unit wavelength in the observed frame, 
 and
 $D_{L}$ is the luminosity distance.
Therefore, 
 we defined the luminosity $L_{4000}$ at rest frame as,
\begin{equation}
\label{eq:l4000}
L_{4000} = (1+z)~4\pi~(D_L(z))^{2}~{\bar{f}_{4000(1+z)}}~,
\end{equation}
 where
 $z$ is the SDSS redshift, 
 ${\bar{f}_{4000(1+z)}}$ is the mean flux 
 at $4000(1+z)\AA$~in the observer frame,
 $D_{L}(z)$ is the luminosity distance. 
The $L_{4000}$ luminosity distribution of these QSOs is
 shown in Figure~\ref{lum_number_rlqrqq}a.
The luminosity distribution of the tQSOs is similar to that of the whole QSO sample:
 both peaks at log($L_{4000})\approx 42.1$.
However,
 the distribution of the rQSOs peaks at log$(L_{4000})\approx 41.1$.
Notably, dust absorption might have led to an underestimation of the luminosity;
therefore, if the redness of these QSOs was caused by dust,
 the QSOs might have lower luminosities.

The luminosity distributions of the
 radio-loud and the radio-quiet sources within the rQSOs and the tQSOs 
 are shown in Figure~\ref{lum_number_rlqrqq}b and Figure~\ref{lum_number_rlqrqq}c, respectively. 
For the tQSOs (Figure~\ref{lum_number_rlqrqq}b),
 the overall luminosity distribution was dominated by the radio-quiet tQSOs
 (hereafter, tRQQs).
The luminosity distributions of the tRQQs and the radio-loud tQSOs
 (hereafter, tRLQs) were similar; 
the distribution peaks of both were at
 log$(L_{4000})\approx 42.1$ and 41.9, respectively.
For the rQSOs (Figure~\ref{lum_number_rlqrqq}c),
 the overall luminosity distribution was also dominated
 by the radio-quiet rQSOs (hereafter, rRQQs).
The distribution peaks for the radio-loud rQSOs (hereafter, rRLQs)
 and rRQQs were both at
 log($L_{4000})\approx  41.1$.
However,  the rRLQs showed a flat distribution,
 which was different from that of the rRQQs.

We showed the number fractions of the rQSOs and the tQSOs 
 at different luminosities  (Figure~\ref{red_fraction}a and Table~\ref{tab_fraction_lum})
 and at different redshifts (Figure~\ref{red_fraction}b and Table~\ref{tab_fraction_z}).
We found that the rQSOs and the tQSOs had different trends 
 along redshifts and luminosities.
The number fractions of the rQSOs decreased with increasing redshifts and luminosities,  
whereas the number fractions of the tQSOs increased with redshifts and luminosities.
We also listed the number fractions of the tRLQs and the rRLQs
 at different luminosities (Table~\ref{tab_fraction_lum_rlq})
 and at different redshifts (Table~\ref{tab_fraction_z_rlq}).
The results shows that at all ranges of luminosities and redshifts,
 the RLQs had a higher number fraction to be red than the whole QSO sample did;
 the overall fraction of the rRLQs in the RLQ group
 ($\xi_{\rm RLQ}(\rm rRLQ)=26.85\%$)
 was approximately twice that of the rQSOs in the whole QSO sample
 ($\xi(\rm rQSO)=13.07\%$). 
We noted that the overall fraction of the tRLQs in the RLQ group
 ($\xi_{\rm RLQ}(\rm tRLQ)=19.14\%$)
 was smaller than the fraction of the tQSOs in the whole QSO sample
 ($\xi(\rm tQSO)=24.90\%$).
These results suggest that the rQSOs are more likely to be radio-loud 
 than the tQSOs.

We also presented the radio-loud fractions of the QSOs
 at different luminosities
 (Figure~\ref{radioloud_fraction}a and Table~\ref{tab_rlq_lum})
 and at different redshifts
 (Figure~\ref{radioloud_fraction}b and Table~\ref{tab_rlq_z}).
Figure~\ref{radioloud_fraction}a shows that
  the radio-loud fractions of the rQSOs and the tQSOs
 increased with luminosities.
Furthermore,
 the rQSOs showed higher radio-loud fractions
 than the tQSOs did 
 at all luminosities.
Moreover, the radio-loud fractions of the rQSOs
 increased with luminosities faster
 than those of the tQSOs.
However, 
 Figure~\ref{radioloud_fraction}b shows that 
 the radio-loud fractions of the tQSOs
  did not change with redshift.
Nevertheless,
 the radio-loud fractions of the rQSOs
 were higher than those of the tQSOs
 at all redshifts.
The radio-loud fractions increased with redshifts at low redshifts ($z = 0.3 - 0.7$),
 but maintained almost no change (or slightly decreased) at high redshifts ($z = 0.8 - 1.1$).
The overall radio-loud fraction of the rQSOs
  ($\xi_{\rm rQSO}(\rm rRLQ)=16.75\%$) 
 was approximately three times that of the tQSOs
  ($\xi_{\rm tQSO}(\rm rRLQ)=6.27\%$). 

Figure~\ref{lum_number_red_norm_3x3} shows 
 the number distribution of the rQSOs and tQSOs at different redshifts.
We found that the number ratios of the rQSOs to the tQSOs
 decreased with increasing redshifts,
 although the number of the tQSOs increased with redshifts.
In addition,
 at high redshifts,
 the rQSOs and the tQSOs seemed to have similar
 luminosity distribution peaks,
 yet at the lowest redshifts ($z = 0.3 - 0.4$),
 the rQSOs had a lower luminosity distribution peak than the tQSOs did.
To investigate whether the red color was due to dust obscuration, 
 we separated the rQSOs into two different color ranges
 with $0.66 \leqslant r< 0.80$ and $r\geqslant 0.80$.
We found that at all redshifts,
 the distributions of these two rQSO groups were
 not significantly different from each other, 
 indicating that the rQSO luminosity is independent of color.

\subsection{\rm Dust -- Luminosity Relation}
To investigate the differences of dust properties between 
 the rQSOs and the tQSOs,
 we compared the dust continuum emission 
 at the rest frame of 10\micron~between these two groups.
Since our QSO samples were at redshifts between 0.3 and 1.2,
 the 10\micron~continuum emission 
 was redshifted to 13\micron~-- 22\micron.
Therefore, 
 we used  the  $M_{10\micron}$ vs. log$(L_{4000})$ diagram 
 (Figure~\ref{lum_10micron}a)
  to present the dust -- luminosity relation for the QSOs,
 where $M_{10\micron}$ is the absolute magnitude
 at the rest frame 10\micron~derived
 from the WISE magnitudes and the SDSS redshifts:
\begin{equation}
\label{eq:micron10}
M_{10\micron} = m_{10\micron} - (5~{\rm log} (D_{\rm L}) -5),
\end{equation}
where $m_{10\micron}$ is the apparent magnitude of the source
 at $10(1+z)\micron$ in the observer frame,
 which can be derived from the interpolation from W3 and W4 magnitude.

Figure~\ref{lum_10micron}a shows that 
 the $M_{10\micron}$~is
 proportional to the log($L_{4000}$).
The fitted lines for the rQSOs and the tQSOs 
 shows a small separation.
The confidence intervals in Figure~\ref{lum_10micron}a and Figure~\ref{lum_10micron}b
 are too small to be easily visible,
 but they are still visible in Figure~\ref{lum_10micron}c.
Since the confidence intervals 
 did not overlap between the rQSO and the tQSO fittings,
the small separation was highly significant.

To determine whether radio activity is related to dust properties, 
 we also present the relation for the RQQs and the RLQs 
 in Figure~\ref{lum_10micron}b and Figure~\ref{lum_10micron}c, respectively.
The results shows that the red samples were more luminous at 10\micron~than the typical samples
 for both the RQQs and the RLQs.
Notably, the infrared magnitude difference
 between the rRLQs and the tRLQs
 was substantially larger than the difference between
 the rRQQs and the tRQQs.

\section{Discussion and Conclusions}
\label{discussion}
Notably, the red QSOs we selected
 might be different from the red QSOs selected for previous studies.
Most red QSOs in previous studies have been
 selected by photometric color cuts
 with spectroscopic follow-ups
 \citep{cut01,gli07,geo09,urr09,gli12,fyn13,ros15}.
Such selections might include different red QSOs 
 depending on different selection cut-offs.
Since our red QSOs were selected based on 
  the flux ratio of the rest frame
  4000\AA~to 3000\AA~continuum emission,
 our data excluded sources
 affected by strong emission lines.
In addition, we selected the red QSOs based on the statistical definition.
Therefore, we could also include slightly-red QSOs,
 which can not be found through photometric selection.

\begin{figure}
\begin{center}
\begin{tabular}{cc}
{\includegraphics[height=8.5cm,angle=-90]{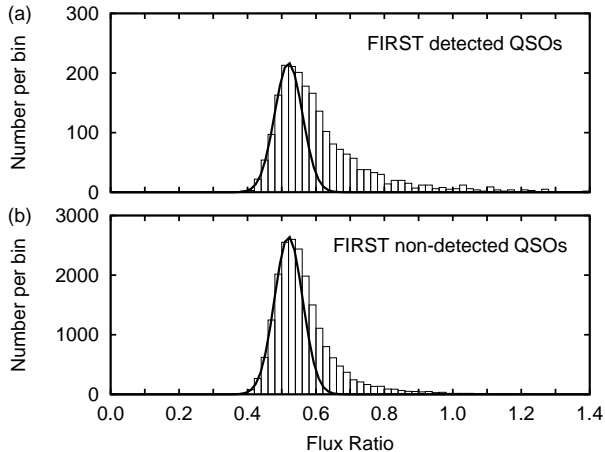}}
\end{tabular}
\end{center}
\caption{
\label{first_det}
{\bf (a)} Number distribution of the QSOs with the FIRST detection at different flux ratios.
{\bf (b)} Number distribution of the QSOs within the FIRST sky coverage but without the FIRST detection.
The number bin width is 0.02.
The black curve in each figure is a Gaussian fitting.
In both figures,  
 the peak centers are at flux ratio $r_{\rm peak} = 0.52 $,
 and $\sigma\approx 0.041$.\\
}
\end{figure}

Since the SDSS QSOs contained both
 UVX-pre-selected and FIRST-pre-selected samples \citep{ric02},
 the color distributions of these two groups of QSOs 
 were expected to be different.
The FIRST-pre-selected QSOs were expected to have a wider
 color distribution  than the UVX-pre-selected QSOs,
 which were expected to be bluer than the FIRST-pre-selected QSOs.
To know whether the properties of our selected red QSOs
 were affected by the selection bias of their FIRST counterparts,
 we plotted the number distribution for
 the radio-detected and the radio-non-detected QSOs
 within the FIRST sky coverage (see Figure~\ref{first_det}).
Both distributions had the appearance of
 a Gaussian with an extended right tail.
We fitted the left wing of the distributions with a Gaussian 
 and folded the result to the right wing.
Both distributions had peaks at $r_{\rm peak}= 0.52$ and $\sigma\approx 0.041$,
 suggesting that the blue wings were similar for both distributions,
 indicating that our red QSO criterion was not affected by
 the FIRST-pre-selection.
We could find red QSOs from the extended red tails 
 whether the QSOs had FIRST detection or not.

However, the red wings were different;
 the red wing for the FIRST-detected QSOs extended further than
 that for the FIRST-non-detected QSOs did.
This result might have been caused by the FIRST-pre-selected QSOs
 having a wider color distribution and more red QSOs.
Notably, the FIRST QSOs were not UVX-pre-selected.
In other words,
 a significant amount of red QSOs might have been missing
 in the UVX-pre-selected process.
These red QSOs were not included in the SDSS catalog
 because of its photometric pre-selection.
Comparing Figure~\ref{first_det}a and~\ref{first_det}b, 
 the results suggest that there should be many more red QSOs,
 which were not found by SDSS because of
 the photometric pre-selection used in SDSS.
In addition, Figure~\ref{radioloud_fraction}b shows that
 the radio-loud fractions of the tQSOs did not change with different redshifts,
 but that the radio-loud fractions of the rQSOs at low redshifts ($z=0.3-0.7$)
 increased with redshifts
 and became almost constant at high redshifts ($z=0.8-1.1$). 
Therefore, explaining the trends of the radio-loud fractions
 at different redshifts with radio selection bias is difficult.
The red QSOs at low and at high redshift might have been caused by different types of reddening,
 or may be the result of selection effects.

Figure~\ref{red_fraction} and Figure~\ref{lum_number_red_norm_3x3}
 also support this argument.
Both figures show that the rQSOs with low luminosities were mainly present at low redshifts,
 but the rQSOs with high luminosities were mainly present at high redshifts.
This could be explained that
 at low redshifts, most rQSOs are dust-obscured tQSOs.
They become faint at high redshifts and are not detectable.
However,
 the rQSOs at high redshifts with high luminosities
 might not be dust-obscured tQSOs,
 but instead might belong to another type of QSOs.

\begin{figure*}
\begin{center}
\begin{tabular}{cc}
{\includegraphics[height=17cm,angle=-90,clip,trim=10.5cm 0cm 0cm 0cm]{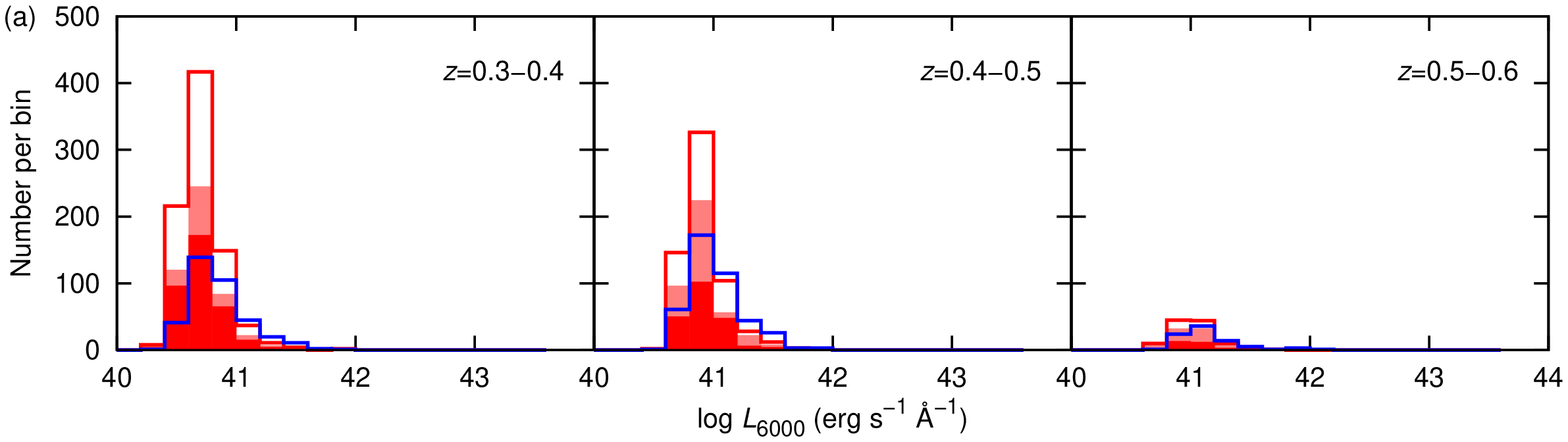}}\\
{\includegraphics[height=17cm,angle=-90,clip,trim=10.5cm 0cm 0cm 0cm]{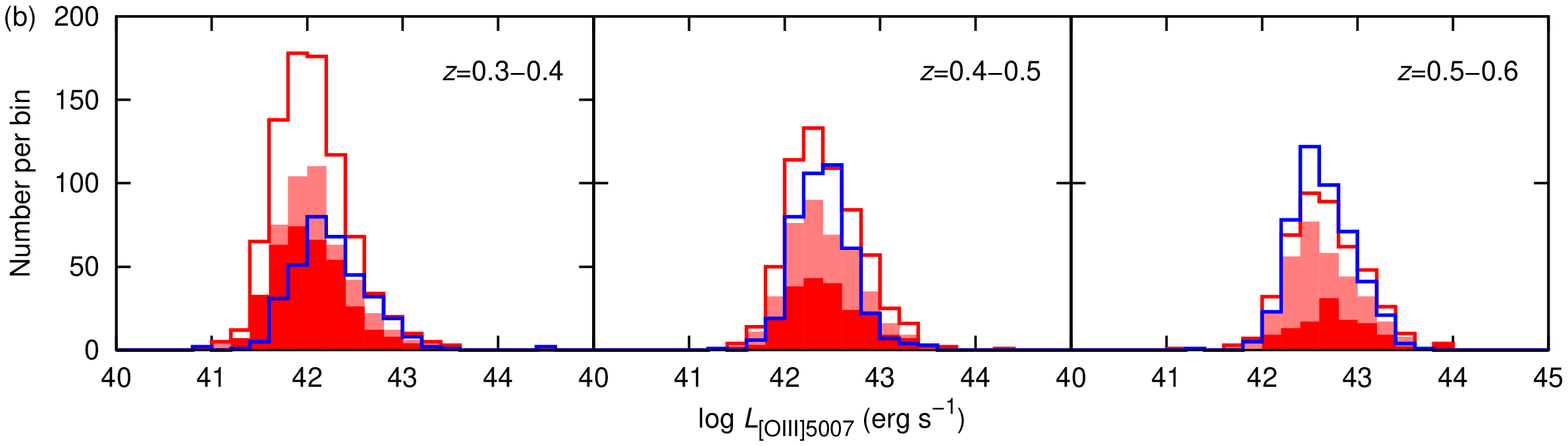}}
\end{tabular}
\end{center}
\caption{
\label{l6000_OIII}
{\bf (a)}
Number distributions of different types of QSOs versus  log($L_{6000}$) at $0.3 <z< 0.6$.
{\bf (b)}
Number distributions of different types of QSOs versus log($L_{\rm [OIII] 5007}$) at $0.3 <z< 0.6$.
The blue solid line represents the tQSOs with flux ratios of
  $0.50 \leqslant r \leqslant 0.54$.
The red solid line represents the rQSOs with flux ratios  $\geqslant 0.66$.
The filled light-red histogram represents the rQSOs with flux ratios of
  $0.66 \leqslant r< 0.80$.
The filled red histogram represents the rQSOs with flux ratios $\geqslant 0.80$.
The values of number, median, and standard deviation for 
 the $L_{\rm 6000}$ and $L_{\rm [OIII] 5007}$ distributions of
 the rQSOs and the tQSOs
 are listed in Table~\ref{stddev_qso}.
The $t$-test results for the 
 $L_{\rm 6000}$ and $L_{\rm [OIII] 5007}$ distributions
 between the rQSOs and the tQSOs
 are listed in Table~\ref{ttest_qso}.
}
\end{figure*}

To investigate whether the red color can be explained 
 solely through dust obscuration  \citep{cut02,gli04,can06},
 we created number distributions at different wavelength (Figure~\ref{l6000_OIII}),
 including the luminosity distribution at 6000\AA~continuum emission (Figure~\ref{l6000_OIII}a),
 and the luminosity distribution of [OIII]~5007\AA~(Figure~\ref{l6000_OIII}b).
We compared these two distributions with that in
 Figure~\ref{lum_number_red_norm_3x3}.
If dust obscuration was the main cause for the red color of the QSOs,
 then absorption would be different at different wavelengths
 following the dust-extinction curve.

Figure~\ref{l6000_OIII}a 
 shows the $L_{6000}$ luminosity distributions
 at low redshifts ($z = 0.3 - 0.6$) 
 because of the limit of the spectrum coverage
 for high-redshift sources.
In Figure~\ref{lum_number_red_norm_3x3} and Table~\ref{stddev_qso},
 the 4000\AA~luminosity distribution at $z = 0.3 - 0.4$ shows that
 the median for the rQSOs is log$(L_{4000}) = 40.73$,
 and that for the tQSOs is log$(L_{4000}) = 41.03$.
The luminosity difference at 4000\AA~is $\Delta$log$(L_{4000}) = 0.3$,
 which would cause a luminosity difference at 6000\AA~of
 $\Delta$log$(L_{6000})\approx 0.2$
 if the difference was caused by the dust absorption
 assuming an SMC dust extinction law \citep{pei92}.
Figure~\ref{l6000_OIII}a and Table~\ref{stddev_qso}
 show that the rQSOs and the tQSOs 
 at $z = 0.3 - 0.4$ had
 a luminosity difference at 6000\AA~of $\Delta$log$(L_{6000})\approx 0.1$.
Furthermore, the rQSOs and the tQSOs
 at $z = 0.4 - 0.6$ had
 a luminosity difference at 6000\AA~of $\Delta$log$(L_{6000}) < 0.1$,
 indicating that dust absorption was not the only cause of redness.

Figure~\ref{l6000_OIII}b shows the luminosity distribution
 of the collisionally excited lines [OIII] 5007\AA~at redshifts of $z = 0.3 - 0.6$.
The [OIII] 5007\AA~is the strongest and least blended narrow emission line in the AGN \citep{bas05}.
The luminosity distribution of [OIII] 5007\AA~depends on AGN activity.
Table~\ref{ttest_qso} shows the $t$-test results
 between the rQSOs and the tQSOs at $z=0.4-0.6$.
It shows that the significance values are $0.3-0.6$ 
 (i.e., the mean values of these two groups are indistinguishable),
 indicating that both groups of QSO at $z=0.4-0.6$
 might show similar AGN activity.
However, the $t$-test results at $z=0.3-0.4$ indicate that
 the rQSOs and the tQSOs have different AGN activities.
The relatively faint luminosity of rQSOs at $z = 0.3 - 0.4$ 
  (Figure~\ref{l6000_OIII}a) might have been related to absorption,
 indicating that the redness of the rQSOs can not be
 explained solely in terms of dust absorption.
Therefore, dust obscuration is not the only cause of the red color for QSOs.

\begin{table*}
\begin{center}
\caption{Number, median, and standard deviation for the luminosity of the QSOs}
\label{stddev_qso}
\begin{tabular}{crrrrrrrr}
\hline
\hline
& 
\multicolumn{2}{c}{4000\AA} & &
\multicolumn{2}{c}{6000\AA} & &
\multicolumn{2}{c}{[OIII]5007\AA} \\
\cline{2-3}\cline{5-6}\cline{8-9}
Redshift & \multicolumn{1}{c}{rQSO}  & \multicolumn{1}{c}{tQSO}
 & & \multicolumn{1}{c}{rQSO}  & \multicolumn{1}{c}{tQSO}
 & & \multicolumn{1}{c}{rQSO}  & \multicolumn{1}{c}{tQSO} \\
\hline
0.3 $\leqslant z <$ 0.4 &  845~(40.73$\pm$0.23) &  365~(41.03$\pm$0.28) & &  844~(40.69$\pm$0.18) &  365~(40.80$\pm$0.25) & &  832~(42.02$\pm$0.39) &  348~(42.21$\pm$0.44) \\
0.4 $\leqslant z <$ 0.5 &  621~(41.04$\pm$0.20) &  426~(41.23$\pm$0.26) & &  621~(40.90$\pm$0.18) &  425~(40.97$\pm$0.23) & &  612~(42.38$\pm$0.38) &  421~(42.39$\pm$0.32) \\
0.5 $\leqslant z <$ 0.6 &  455~(41.26$\pm$0.22) &  476~(41.42$\pm$0.26) & &  113~(41.00$\pm$0.17) &   83~(41.07$\pm$0.26) & &  447~(42.63$\pm$0.40) &  466~(42.61$\pm$0.32) \\
0.6 $\leqslant z <$ 0.7 &  282~(41.48$\pm$0.21) &  548~(41.59$\pm$0.24) & &    3~(41.39$\pm$0.29) &    1~(41.61$\pm$0.00) & &  280~(42.79$\pm$0.40) &  532~(42.78$\pm$0.33) \\
0.7 $\leqslant z <$ 0.8 &  160~(41.71$\pm$0.23) &  594~(41.78$\pm$0.24) & &    1~(41.81$\pm$0.00) & \multicolumn{1}{c}{-} & &  156~(43.04$\pm$0.36) &  585~(43.00$\pm$0.35) \\
0.8 $\leqslant z <$ 0.9 &  158~(41.85$\pm$0.20) &  636~(41.89$\pm$0.22) & & \multicolumn{1}{c}{-} & \multicolumn{1}{c}{-} & &   51~(43.27$\pm$0.40) &  161~(43.20$\pm$0.31) \\
0.9 $\leqslant z <$ 1.0 &  140~(42.02$\pm$0.20) &  700~(42.00$\pm$0.23) & & \multicolumn{1}{c}{-} & \multicolumn{1}{c}{-} & & \multicolumn{1}{c}{-} & \multicolumn{1}{c}{-} \\
1.0 $\leqslant z <$ 1.1 &  144~(42.14$\pm$0.19) &  841~(42.09$\pm$0.22) & & \multicolumn{1}{c}{-} & \multicolumn{1}{c}{-} & & \multicolumn{1}{c}{-} &    1~(43.43$\pm$0.00) \\
1.1 $\leqslant z <$ 1.2 &  126~(42.25$\pm$0.20) &  996~(42.18$\pm$0.21) & & \multicolumn{1}{c}{-} & \multicolumn{1}{c}{-} & & \multicolumn{1}{c}{-} & \multicolumn{1}{c}{-} \\
\hline
\multicolumn{9}{l}{
\begin{minipage}[t]{2\columnwidth}
\begin{enumerate}
\itemsep 0em
\item
The integer value is the number of different types of QSOs 
 within the corresponding redshift range. 
\item
The values inside the parenthesis are the median and the standard deviation of
 logarithmic scale luminosity in different types of QSOs 
 within the corresponding redshift range. 
\item
The unit of luminosity for both 4000\AA~and 6000\AA~is erg s$^{-1}$\AA$^{-1}$, and that for [OIII]5007\AA~is erg s$^{-1}$.
\item
The symbol ``-'' means there is no QSO found.
\item
The standard deviation of $\pm$0.00 means there is only one QSO found.
\end{enumerate}
\end{minipage}
}\\
\end{tabular}
\end{center}
\end{table*}

\begin{table}
\begin{center}
\caption{The $t$-test of $L_{\rm [OIII]5007}$ between the rQSOs and the tQSOs}
\label{ttest_qso}
\begin{tabular}{crrc}
\hline
\hline
 Redshift & & 
\multicolumn{1}{c}{$t$}  &  \multicolumn{1}{c}{$s$}  \\
\hline
 0.3 $\leqslant z <$ 0.4 & &  -7.9907 &   0.0000  \\
 0.4 $\leqslant z <$ 0.5 & &   0.5531 &   0.5803  \\
 0.5 $\leqslant z <$ 0.6 & &   0.9368 &   0.3491  \\
\hline
\multicolumn{4}{l}{
\begin{minipage}[t]{0.7\columnwidth}
\begin{enumerate}
\itemsep 0em
\item
$t$ is the $t$-test statistic.
\item
$s$ is the significance. The significance is a value in the
interval [0.0, 1.0]. A small value ($s<0.01$) indicates
that two samples have significantly different means.
\end{enumerate}
\end{minipage}
}\\
\end{tabular}
\end{center}
\end{table}

Figure~\ref{lum_10micron} also supports this argument.
From the dust -- luminosity relation,
 we found that the rQSOs had a stronger 10\micron~emission than the tQSOs did,
 implying that the rQSOs might have had more hot dust than the tQSOs had.
Furthermore, the infrared magnitude differences in the RLQs were
 substantially larger than those in the RQQs.
Figure~\ref{lum_10micron}c shows that
 for a given absolute magnitude at 10\micron,
 the magnitude difference between the rRLQs
 and the tRLQs was approximately 0.5 magnitude,
 which was larger than the magnitude difference 
 subject to by dust obscuration of $<0.26$ magnitude (see Section~2.1).
In other words, 
 the influence of the assumed dust extinction
 is not as major a cause as the results suggest.

Moreover, Figure~\ref{lum_number_red_norm_3x3} shows
 that the luminosity peaks between
 the rQSOs and the tQSOs 
 showed no difference in most redshifts ($z = 0.4 - 1.1$)
 except for the lowest redshifts ($z = 0.3 - 0.4$)
 where they were significantly different.
If the rQSOs were tQSOs with dust obscuration,
 then the rQSOs should have had a lower luminosity
 than the tQSOs had at all redshifts.
However, the luminosity distributions 
 do not support the dust obscuration scenario.
Furthermore, if a redder color represented more dust absorption, 
 then the two groups of rQSOs 
(less red color with $0.66\leqslant r < 0.80$
  and much red color with $r\geqslant 0.80$)
 in Figure~\ref{lum_number_red_norm_3x3} 
 should show different luminosity peaks.
However, the result does not show a significant difference 
 between these two red groups:
 the luminosities of the rQSOs were independent of their colors,
 which also reveals that 
 most rQSOs in this study were not from dust-obscured sources.

On the other hand,
 the luminosity distributions of the rQSOs at low redshifts
 were different from those at high redshifts.
At high redshifts ($z = 0.6 - 1.2$),
 the number ratios of the rQSOs to the tQSOs were small,
 whereas at low redshifts ($z = 0.3 - 0.6$),
 the number ratios of the rQSOs to the tQSOs were large.
In addition, most rQSOs at low redshifts
 had relatively lower luminosities than the tQSOs had.
In other words, the observed red QSOs showed extremely
 different luminosity distributions at different
 redshifts, and might have had different origins.

\begin{figure*}
\begin{center}
\begin{tabular}{cc}
{\includegraphics[height=17cm,angle=-90,clip,trim=5cm 0cm 0cm 0cm]{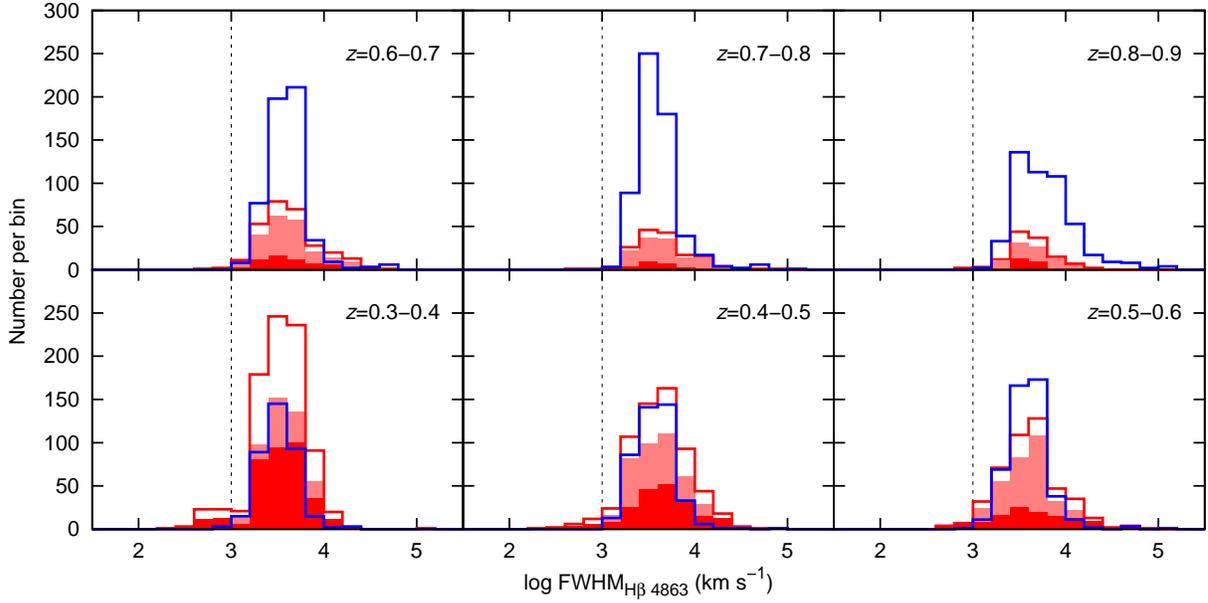}}
\end{tabular}
\end{center}
\caption{
\label{Hbeta_4863}
Number distributions of different types of QSOs versus log(FWHM$_{\rm H\beta~4863}$) at $0.3 <z< 0.9$.
The log(FWHM$_{\rm H\beta~4863}$) is the Gaussian-fit FWHM
 of the H$\beta$ emission line  obtained from the SDSS database.
The blue solid line represents the tQSOs with flux ratios of
  $0.50 \leqslant r \leqslant 0.54$.
The red solid line represents the rQSOs with flux ratios  $\geqslant 0.66$.
The filled light-red histogram represents the rQSOs with flux ratios of
  $0.66 \leqslant r< 0.80$.
The filled red histogram represents the rQSOs with flux ratios $\geqslant 0.80$.
The dotted line represents the H$\beta$ FWHM of 1000~km~s$^{-1}$.
The number of QSOs with the H$\beta$ FWHM  $<$1000~km~s$^{-1}$ is listed in Table~\ref{vel999}.
\\
}
\end{figure*}

We also investigated whether the redness of the QSOs were caused by type II QSOs
 in the SDSS QSO samples.
Some QSOs in the SDSS QSO catalog might have been from
 narrow-line sources \citep{sch10}.
We plotted the FWHM distribution 
 of the H$\beta$ 4863\AA~emission for the rQSOs and the tQSOs
 (Figure~\ref{Hbeta_4863}),
 and found that most QSOs were broad-line sources;
 only a few were narrow-line sources,
 especially among the rQSOs at low redshifts with low luminosities.
Approximately 3.6\% of the rQSOs were narrow-line sources,
 and 0.14\% of the tQSOs were narrow-line sources (see Table~\ref{vel999}).
These sources might have belonged to type II QSOs and been caused by dust absorption,
 or might be contaminated by the red light of stars in their host galaxies.
However, the redness of the broad-line rQSOs might have different origins.
These results support that there were at least two types of
 red QSOs in our red sample.

Notably,
 the colors of our rQSOs were independent of redshifts (Figure~\ref{fluxratio_z}).
Since our rQSOs were selected from the QSO samples
 with color deviation $>3\sigma$,
 the number of rQSOs were $<0.3\%$ of the QSO samples
 when the rQSOs followed a normal distribution.
However,
 the fractions of the rQSOs in the whole QSO sample at low redshifts
 (e.g., $\xi({\rm rQSO)}=25.95-34.55\%$ at $z = 0.3 - 0.5$
 as shown in Table~\ref{tab_fraction_z})
 was approximately $2 - 3$ times larger than
 the overall fraction of the rQSOs in the whole QSO sample
 ($\xi({\rm rQSO})=13.07\%$).
The fractions of the rQSOs in the whole QSO sample
 at high redshifts could be much higher than we originally expected.

In conclusion, 
 we suggest that at least two types of red QSOs 
 are present in our red sample.
One group are the bright red QSOs,
 which have a luminosity distribution similar to that of the typical QSOs at the same redshifts.
The number of bright red QSOs
 is substantially lower than that of typical QSOs.
The red color 
 is not due dust-obscured sources from typical QSOs.
The other group is the faint red QSOs,
 which have lower luminosities than those of the typical QSOs at the same redshifts.
Here the red color could be related to dust obscuration.
Notably,
 these faint red QSOs can easily be detected at low redshifts,
 whereas at high redshifts,
 they cannot be detected because of the detection limit.

\begin{table}
\begin{center}
\caption{Number and fractions of QSOs with the H$\beta$ FWHM $<$1000~km~s$^{-1}$}
\label{vel999}
\begin{tabular}{cccc}
\hline
\hline
Redshift & QSOs &  rQSOs  & tQSOs \\
\hline
0.3 $\leqslant z <$ 0.4   & \enspace77 (3.15\%) &        50 (5.92\%) &   3 (0.82\%) \\
0.4 $\leqslant z <$ 0.5   & \enspace24 (1.00\%) &        22 (3.54\%) &   0 (0.00\%) \\
0.5 $\leqslant z <$ 0.6   & \enspace14 (0.59\%) &        11 (2.43\%) &   1 (0.21\%) \\
0.6 $\leqslant z <$ 0.7   &     \quad5 (0.22\%) & \enspace3 (1.07\%) &   0 (0.00\%) \\
0.7 $\leqslant z <$ 0.8   &     \quad3 (0.14\%) & \enspace2 (1.25\%) &   0 (0.00\%) \\
0.8 $\leqslant z <$ 0.9   &     \quad2 (0.11\%) & \enspace2 (1.64\%) &   0 (0.00\%) \\
0.9 $\leqslant z <$ 1.0   &     \quad0 (0.00\%) & \enspace0 (0.00\%) &   0 (0.00\%) \\
1.0 $\leqslant z <$ 1.1   &     \quad0 (0.00\%) & \enspace0 (0.00\%) &   0 (0.00\%) \\
1.1 $\leqslant z <$ 1.2   &     \quad0 (0.00\%) & \enspace0 (0.00\%) &   0 (0.00\%) \\
\hline
0.3 $\leqslant z <$ 1.2   &   125 (0.93\%) &  90 (3.63\%) &   4 (0.14\%) \\
\hline
\multicolumn{4}{l}{
\begin{minipage}[t]{0.85\columnwidth}
\begin{enumerate}
\itemsep 0em
\item
The integer value is the number of different types of QSOs
 within the corresponding redshift range.
\item
The value inside the parenthesis is the number fraction
 of different types of QSOs
 within the corresponding redshift range.
\end{enumerate}
\end{minipage}
}\\
\end{tabular}
\end{center}
\end{table}

\begin{comment}
\begin{table}
\begin{center}
\caption{Number of QSOs with the H$\beta$ FWHM $<$1000~km~s$^{-1}$}
\label{vel999}
\begin{tabular}{crrrrrr}
\hline
\hline
Redshift &  rQSO  & tQSO \\
\hline
0.3 $\leqslant z <$ 0.4 &  50 &   3 \\
0.4 $\leqslant z <$ 0.5 &  22 &   0 \\
0.5 $\leqslant z <$ 0.6 &  11 &   1 \\
0.6 $\leqslant z <$ 0.7 &   3 &   0 \\
0.7 $\leqslant z <$ 0.8 &   2 &   0 \\
0.8 $\leqslant z <$ 0.9 &   2 &   0 \\
0.9 $\leqslant z <$ 1.0 &   0 &   0 \\
1.0 $\leqslant z <$ 1.1 &   0 &   0 \\
1.1 $\leqslant z <$ 1.2 &   0 &   0 \\
\hline
\end{tabular}
\end{center}
\end{table}
\end{comment}

\bigskip
\begin{acknowledgements}
We are appreciative of Anthony Moraghan for English correction. 
We also thank the anonymous referee for very useful comments.

This work is supported by the Ministry of Science and Technology (MOST) of Taiwan,
MOST 103-2119-M-008-017-MY3.

This publication makes use of data products from the Wide-field Infrared Survey Explorer, which is a joint project of the University of California, Los Angeles, and the Jet Propulsion Laboratory/California Institute of Technology, funded by the National Aeronautics and Space Administration.

Funding for SDSS-III has been provided by the Alfred P. Sloan Foundation, the Participating Institutions, the National Science Foundation, and the U.S. Department of Energy Office of Science. The SDSS-III web site is http://www.sdss3.org/.

SDSS-III is managed by the Astrophysical Research Consortium for the Participating Institutions of the SDSS-III Collaboration including the University of Arizona, the Brazilian Participation Group, Brookhaven National Laboratory, Carnegie Mellon University, University of Florida, the French Participation Group, the German Participation Group, Harvard University, the Instituto de Astrofisica de Canarias, the Michigan State/Notre Dame/JINA Participation Group, Johns Hopkins University, Lawrence Berkeley National Laboratory, Max Planck Institute for Astrophysics, Max Planck Institute for Extraterrestrial Physics, New Mexico State University, New York University, Ohio State University, Pennsylvania State University, University of Portsmouth, Princeton University, the Spanish Participation Group, University of Tokyo, University of Utah, Vanderbilt University, University of Virginia, University of Washington, and Yale University.

\end{acknowledgements}
\bigskip

\bibliographystyle{apj}

\clearpage

\end{document}